# Faster SDP hierarchy solvers for local rounding algorithms[*]

Venkatesan Guruswami[†]    Ali Kemal Sinop[‡]

November 3, 2018

## Abstract

Convex relaxations based on different hierarchies of linear/semi-definite programs have been used recently to devise approximation algorithms for various optimization problems. The approximation guarantee of these algorithms improves with the number of *rounds* $r$ in the hierarchy, though the complexity of solving (or even writing down the solution for) the $r$'th level program grows as $n^{\Omega(r)}$ where $n$ is the input size.

In this work, we observe that many of these algorithms are based on *local* rounding procedures that only use a small part of the SDP solution (of size $n^{O(1)}2^{O(r)}$ instead of $n^{\Omega(r)}$). We give an algorithm to find the requisite portion in time polynomial in its size. The challenge in achieving this is that the required portion of the solution is not fixed a priori but depends on other parts of the solution, sometimes in a complicated iterative manner.

Our solver leads to $n^{O(1)}2^{O(r)}$ time algorithms to obtain the same guarantees in many cases as the earlier $n^{O(r)}$ time algorithms based on $r$ rounds of the Lasserre hierarchy. In particular, guarantees based on $O(\log n)$ rounds can be realized in polynomial time. For instance, one can (i) get $O(1/\lambda_r)$ approximations for graph partitioning problems such as minimum bisection and small set expansion in $n^{O(1)}2^{O(r)}$ time, where $\lambda_r$ is the $r$'th smallest eigenvalue of the graph's normalized Laplacian; (ii) a similar guarantee in $n^{O(1)}k^{O(r)}$ for Unique Games where $k$ is the number of labels (the polynomial dependence on $k$ is new); and (iii) find an independent set of size $\Omega(n)$ in 3-colorable graphs in $(n2^r)^{O(1)}$ time provided $\lambda_{n-r} < 17/16$.

We develop and describe our algorithm in a fairly general abstract framework. The main technical tool in our work, which might be of independent interest in convex optimization, is an efficient ellipsoid algorithm based separation oracle for convex programs that can output a *certificate of infeasibility with restricted support*. This is used in a recursive manner to find a sequence of consistent points in nested convex bodies that "fools" local rounding algorithms.

[*]An extended abstract of this work appears in the *Proceedings of FOCS 2012*. This is an expanded version.

[†]Computer Science Department, Carnegie Mellon University. Supported in part by NSF grants CCF-0963975 and CCF-1115525. Email: `guruswami@cmu.edu`

[‡]Center for Computational Intractability, Department of Computer Science, Princeton University. Work was done while the author was a student at Carnegie Mellon University and was supported by NSF CCF-1115525, and MSR-CMU Center for Computational Thinking. Email: `asinop@cs.cmu.edu`

# Contents

# List of Figures

# List of Algorithms

# 1 Introduction

A rich body of recent research has shown that for many optimization problems, the Unique Games conjecture (UGC) serves as a barrier to further improvements to the approximation factor achieved by efficient algorithms. In many cases, including all constraint satisfaction problems and various graph partitioning problems, the best algorithms are based on fairly simple semi-definite programming (SDP) relaxations. The UGC foretells that for these problems, no tighter relaxation than these simple SDPs will yield a better approximation ratio in the worst-case.

**Hierarchies of convex relaxations.** A natural question thus is to understand the power and limitations of potentially stronger SDP relaxations, for example those from various *hierarchies* of relaxations. These hierarchies are parameterized by an integer $r$ (called rounds/levels) which capture higher order correlations between (roughly $r$-tuples of) variables (the basic SDP captures only pairwise correlations, and certain extensions like triangle inequalities pose constraints on triples). Larger the $r$, tighter the relaxation. The optimum of $n$'th level of the hierarchy, where $n$ is the number of variables in the underlying integer program, usually equals the integral optimum.

There are several hierarchies of relaxations that have been studied in the literature, such as Sherali-Adams hierarchy of linear programs [SA90], the Lovász-Schrijver hierarchy [LS91], a "mixed" hierarchy combining Sherali-Adams linear programs with the base level SDP, and the Lasserre hierarchy [Las02] (see [CT11] for a recent survey focusing on their use in approximate combinatorial optimization). Of these hierarchies, the most powerful one is the Lasserre hierarchy (see [Lau03] for a comparison), and therefore holds the most potential for new breakthroughs in approximation algorithms. Arguably, Lasserre SDPs pose the currently strongest known threat to the Unique Games conjecture, as even the possibility of the 4'th level of Lasserre SDP relaxation improving upon the Goemans-Williamson 0.878 approximation factor for Max Cut has *not* been ruled out. Recently, it has also been shown that $O(1)$ rounds of the Lasserre hierarchy are able to solve all candidate gap instances of Unique Games [BHK+12]. (On the other hand, for some of the weaker hierarchies, integrality gaps for super-constant rounds *are* known for various Unique-Games hard problems [KS09, RS09].)

In light of the above, the power and limitations of the Lasserre hierarchy merit further investigation. There has been a fair bit of recent interest in Lasserre hierarchy based approximation algorithms [CS08, KMN10, GS11, BRS11, RT12, AG11, GS12a]. For instance, our work [GS11] shows that various graph partitioning problems (including minimum bisection, sparsest cut, and Unique Games) can be well-approximated by $\approx r$ rounds of the Lasserre SDP on graphs whose $r$'th smallest eigenvalue $\lambda_r$ (of the Laplacian) is reasonably large.

A (near)-optimal solution to the $r$'th level Lasserre relaxation can be found in $n^{O(r)}$ time. So understanding the power of these relaxations for small values of $r$ is of particular interest. The main contribution of this work is to improve the running time of various Lasserre-based approximation algorithms to $2^{O(r)}n^{O(1)}$ (from the default $n^{O(r)}$). In particular, the guarantees achieved by $O(\log n)$ rounds of Lasserre SDPs can be realized in polynomial time. Plugging our methods into some of the algorithms in [GS11] gives us the following application:

**Theorem 1.** *For the graph partitioning problems such as uniform sparsest cut, small set expansion, and minimum bisection, one can compute a cut with cost at most $\frac{1.5}{\min\{1,\lambda_r\}}$ times the optimum in $n^{O(1)}2^{O(r)}$ time, where $n$ is the number of vertices and $\lambda_r$ the $r$'th smallest eigenvalue of the normalized Laplacian of the input graph.[1]*

---
[1]For the case of minimum bisection, the cut may have a $o(n)$ imbalance. In fact, an approximation factor of

*For Unique Games with k labels and n variables, an approximation factor of $3/\min\{1, \lambda_r\}$ can be achieved for the minimization version in time $n^{O(1)}k^{O(r)}$, where $\lambda_r$ the r'th smallest eigenvalue of the normalized Laplacian of the constraint graph. (Note the polynomial dependence of the runtime on the number of labels, which is new to this work.)*

Our methods also apply to Lasserre hierarchy based approximation algorithms discovered recently. A table of several algorithms whose algorithms we are able to improve is given in Section 7. In particular, we describe how Lasserre hierarchy based algorithms for finding independent sets in 3-colorable graphs [AG11] and approximating 2-CSPs [BRS11] can be fit into our framework. The details of this are described in Appendix B, and along the way, in Appendix A, we collect useful geometric bounds on the variance of conditioned solutions in Lasserre SDPs. As a result, we get the following speed-ups to the algorithms in [BRS11, AG11].

**Theorem 2** (Informal). *Given a 3-colorable regular graph $G$, we can find an independent set of size at least $n/12$ in $2^{O(r)}n^{O(1)}$ time when the r'th largest eigenvalue of the normalized Laplacian of $G$ is at most $17/16$.*

**Theorem 3** (Informal). *Let $\mathcal{P}$ be an arity 2 CSP over a domain of size $k$, and let $\varepsilon \in (0, 1)$. Given an instance of $\mathcal{P}$ whose constraint graph $G$ is regular, we can find an assignment satisfying at least $\mathsf{opt}(G) - \varepsilon$ fraction of constraints in $2^{r \operatorname{poly}(k/\varepsilon)} n^{\operatorname{poly}(k/\varepsilon)}$ time assuming that the r'th smallest eigenvalue of the normalized Laplacian of $G$ is at least $1 - \frac{\varepsilon^2}{2k^2}$. (Here $\mathsf{opt}(G)$ denotes the fraction of constraints of $G$ satisfied by an optimal assignment.)*

Our techniques might also be useful in the context of fixed-parameter tractability, which we leave as a potentially interesting avenue for future research.

**Local rounding algorithms.** Note that even writing down the full $r$-round Lasserre solution takes $n^{\Omega(r)}$ time. The hope to speed-up the algorithms to a runtime dependence of $2^{O(r)}$ is based on the observation that many of the rounding algorithms have a "local" character that uses only a small portion of the SDP solution. In the simplest setting, the rounding algorithm proceeds in two steps: (i) find a "seed set" $S^*$ of $\approx r$ nodes based only the solution to the base (1-round) SDP, and (ii) use the value of $r$-round Lasserre solution on the set $S^*$ to sample a partial assignment to $S^*$ and then propagate it to the other nodes. Thus the rounding algorithm only uses the portion of the Lasserre SDP corresponding to the subsets $S^* \cup \{u\}$ for various $u$. Further, the analysis of the rounding algorithm also relies only on Lasserre consistency constraints for subsets of $S^*$. More generally, the algorithms might pick a sequence of seed subsets $S_1, S_2, \ldots, S_\ell$ iteratively and the SDP solution restricted to subsets of $S_1 \cup S_2 \cup \cdots \cup S_\ell$ is used for rounding.

Note that the needed portion of the solution (corresponding to $S^*$, or more generally $S_1 \cup S_2 \cup \cdots S_\ell$) itself depends on certain other parts of the solution. So one cannot simply project the space down to the relevant dimensions to find the required part of the Lasserre solution. Our main technical contribution is an ellipsoid algorithm that can find the needed partial solution (which satisfies all the local constraints induced on those variables) in time polynomial in the number of variables in the partial solution. We stress that the partial solution we find *may not* extend to a full Lasserre solution. This, however, does not matter for the approximation guarantee as it will *"fool" the rounding algorithm* which can't distinguish the solution we find from a global Lasserre solution.

There are two examples of hierarchy based approximation algorithms which have been sped up to $2^{O(r)}$ dependence on the number of rounds, both of which rely on weaker hierarchies than

---

$\frac{1+\varepsilon}{\min\{1,\lambda_r\}}$ can be achieved in $n^{O(1/\varepsilon^2)} 2^{O(r/\varepsilon)}$ time.

Lasserre: (i) the algorithm for sparsest cut on bounded treewidth graphs using the Sherali-Adams hierarchy [CKR10] and (ii) the Unique Games algorithm based on the "mixed" hierarchy [BRS11]. The faster algorithm is for the former case is immediate as the required portion of the solution only depends on the input graph, so one can simply find that part using any LP solver. For the Unique Games algorithm, the seed set $S^*$ depends on the vector solution to the basic SDP relation. The goal is to extend the solution to local distributions of labels on subsets $S^* \cup \{u\}$ for various nodes $u$, whose 2-way marginals agree with the vector inner products. As briefly sketched in [BRS11], these constraints form a linear program, and if infeasible, by Farkas' lemma, one can get a new constraint for the vector inner products, which can be fed into an ellipsoid algorithm for solving the basic SDP. Our situation is more complicated as we handle several iterations of seed set selection, and the "extension" problems we solve are no longer simple linear programs. Also, the runtime of the Unique Games algorithm in [BRS11] had an exponential dependence on the number of labels, as opposed to our polynomial dependence.

**Main technique: Separation oracle with restricted support.** We describe the high level ideas behind our method for finding adequate partial solutions to Lasserre SDPs in Section 2. Our approach applies in a fairly general set-up, and therefore we describe our methods in an abstract framework for clarity, both in Section 2 and later in Section 4 where the formal details appear.

In addition to the runtime improvements, our results contribute a useful, and to our knowledge new, basic tool in convex optimization, which is an ***efficient ellipsoid algorithm based separation oracle that can output a certificate of infeasibility with restricted support*** (or more generally belonging to a restricted subspace). For instance, suppose we are given a convex body $K \subseteq \mathbb{R}^n$ via a separation oracle for it. Given a point $y \in \mathbb{R}^U$ (a potential partial solution) for some $U \subset \{1, 2, \ldots, n\}$, we give an algorithm to either find $x \in K$ such that $\text{proj}_U(x) = y$ (if one exists[2]), or find a separating constraint *that is supported on $U$*.

## 1.1 Organization

Our paper is organized in the following way. In Section 2, we formalize the notion of local rounding algorithms and briefly introduce our main technical contribution. We then present some mathematical preliminaries in Section 3 and our main technical contribution in Section 4. Our final solver is given in Section 5.

In Section 6, we introduce Lasserre Hierarchy in a way suitable for our framework. Finally we state the faster running times we obtain for various rounding algorithms based on Lasserre Hierarchy in Section 7.

For the semi-coloring algorithm from [AG11] and the 2-CSP algorithm from [BRS11], we present the faster rounding algorithms and prove their correctness in Appendix B. Along the way, in Appendix A, we give a geometric interpretation of variance reduction via conditioning as needed in [BRS11, AG11], using terminology and methods from [GS11]. We believe this could be useful elsewhere.

---
[2] Actually, we need the volume of $K \cap \text{proj}_U^{-1}(y)$ to be at least some small $\varepsilon$

# 2 An abstract framework of local rounding and Overview of our techniques

Consider a rounding algorithm with following property: Given an optimal solution $x \in \mathbb{R}^N$ as input, it only reads a much smaller part of this solution, say $T \subseteq N$ with $|T| = o(|N|)$. We call these "local" rounding algorithms: Even though this setting might sound too restrictive and/or unrealistic, observe that several of the known rounding algorithms which use "hierarchies" fit into this framework, [CS08, KMN10, GS11, AG11, RT12, GS12a]. See Section 7 for details.

**Local Rounding.** We first start by outlining a generic *iterative rounding* algorithm. This framework depends on two application specific deterministic[3] procedures, SEED and FEASIBLE. Without going into the formal details, at a high level, $\mathsf{SEED}_S$ procedure chooses next "seed set" designating which fragment of the solution we will read based on current seeds $S$ and $\mathsf{FEASIBLE}_S(y)$ is a *strong separation oracle* for a convex body $K_S$ representing the induced solutions on seeds $S$.

At the end, final seeds and induced solution are fed into another application specific rounding procedure.

**Formal Framework.** Given two problem specific procedures, SEED and FEASIBLE, we formalize the generic algorithm described above as follows.

1. Let $x \in \mathbb{R}^N$ be a vector representing an optimal solution for some convex optimization problem, $x \in K_N$.
2. Let $S(0)$ be the initial solution fragment and $y(0) \leftarrow x_{S(0)}$ be the induced solution.
3. For $i \leftarrow 0$ to $\ell$:
   (a) Fail if $\mathsf{FEASIBLE}_{S(i)}(y(i))$ asserts infeasible (i.e. $y(i) \notin K_{S(i)}$).
   (b) If $i < \ell$, read next part of solution: $S(i+1) \leftarrow \mathsf{SEED}_{S(i)}(y(i))$ and $y(i+1) \leftarrow x_{S(i+1)}$.
4. Perform rounding using $S(\ell)$ and $y(\ell)$.

**Our Goal.** Suppose $|S(\ell)| \ll |N|$ – the algorithm reads only a negligible portion of the full solution. Then can we find an equivalent rounding algorithm which runs in time $\mathrm{poly}(|S(\ell)|)$ as opposed to $\mathrm{poly}(N)$? Claim 4 shows this can be expected:

**Claim 4.** *Above rounding algorithm can not distinguish between the following two cases, i.e. any properties satisfied by the output assuming 1 still holds under a weaker condition, 2:*

1. *There exists a feasible solution $x \in \mathbb{R}^N$, i.e. $\mathsf{FEASIBLE}_N(x)$ asserts feasible.*
2. *For all $i \in \{0, \ldots, \ell\}$:*
   - *$y(i) \in K_{S(i)}$: $\mathsf{FEASIBLE}_{S(i)}(y(i))$ asserts feasible,*
   - *$S(i+1) = \mathsf{SEED}_{S(i)}(y(i))$ if $i < \ell$,*
   - *$y(i+1)_{S(i)} = y(i)$ if $i < \ell$.*

Using this insight, we first consider a simple case and give an algorithm whose running time depends on $|S(\ell)|$ instead of $|N|$.

---
[3]We can allow randomization also, but we stick to the deterministic case for simplicity, since all the seed selection procedures used by the known algorithms can be derandomized.

## 2.1 An Algorithm for a Simple Case

Suppose that SEED procedure does not depend on $y$. Then the above conditions can easily be expressed as a convex problem of size $|S(\ell)|$, which is much smaller than the original problem. Then we can solve this convex problem using standard ellipsoid procedure and execute the above procedure on this solution instead.

## 2.2 Our Algorithm

Unfortunately for all algorithms we consider in this paper, the procedure SEED heavily depends on $y$. In particular, at the $i^{th}$ level, $0 \leqslant i < \ell$, we are trying to solve the following induced problem on $S(i+1)$. Given $y(i) \in K_{S(i)}$:

$$
\begin{aligned}
\text{Find} \quad & y(i+1) \\
\text{st} \quad & y(i+1)_{S(i)} = y(i),\ y(i+1) \in K_{S(i+1)}; \\
& \exists y(i+2) \in K_{S(i+1)} : y(i+2)_{S(i+1)} = y(i+1) \\
& \qquad \text{where } S(i+2) = \mathsf{SEED}_{S(i+1)}(y(i+1)); \\
& \vdots \\
& \exists y(\ell) \in K_{S(\ell)} : y(\ell)_{S(\ell-1)} = y(\ell-1) \\
& \qquad \text{where } S(\ell) = \mathsf{SEED}_{S(\ell-1)}(y(\ell-1)).
\end{aligned}
\tag{1}
$$

Observe that if we can construct a weak separation oracle for eq. (1) at $(i+1)^{th}$ level, then we can combine it with ellipsoid algorithm to solve the problem at $i^{th}$ level also. Thus if we can convert this ellipsoid algorithm to a weak separation oracle, then we can call these separation oracles recursively starting from $0^{th}$ level all the way down to $\ell^{th}$ level:

**Recursive Separation Oracle.** (Template for $i^{th}$ level)

1. Given $S(i)$ and $y(i)$, if $\mathsf{FEASIBLE}_{S(i)}(y(i))$ asserts infeasible and returns $c$, then assert infeasible and return $c$ (to the $(i-1)^{th}$ level).
2. If $i = \ell$, then return the solution $y(\ell)$.
3. Let $S(i+1) \leftarrow \mathsf{SEED}_{S(i)}(y(i))$.
4. Use ellipsoid method to find $y(i+1)$ such that $y(i+1)_j = y(i)_j$ for all $j \in S(i)$ with separation oracle being a recursive call for the $(i+1)^{th}$ level (which takes inputs $S(i+1)$ and $y(i+1)$).
5. If ellipsoid method fails to find such solution $y(i+1)$, return a separating hyperplane.

The key question now is how one might implement (the currently vague) step 5. Let us inspect a simple option, and see what goes wrong with it.

**Return an arbitrary hyperplane seen so far.** Any inequality returned by the recursive separation oracle call is a valid separating hyperplane, so consider returning an arbitrary one. What goes wrong in this case? The problem is that the running time now might be as large as polynomial in $|N|$. To see this, suppose that $\mathsf{FEASIBLE}_{S(\ell)}(y(\ell))$ returned an inequality on support $S(\ell)$. Then the parent ellipsoid procedure needs to keep track of the additional variables from this particular $S(\ell)$, call it $\tilde{S}$. At some later stage, the algorithm may backtrack and change an earlier seed set, say $S(\ell - 5)$, which will need to a new $S(\ell)$. But the algorithm would still need to keep the values of variables from the $\tilde{S}$, the old value of $S(\ell)$. Continuing in this fashion, the set of variables the algorithm has to track might end up being $N$, which is equivalent to constructing the whole solution on $\mathbb{R}^N$!

This attempt has not been futile though, as it shows what kind of hyperplanes we need:

$$\text{Any hyperplane returned by step 5 at } (i+1)^{st} \text{ level should have support } S(i). \quad (2)$$

We outline our proposed solution in the next section.

### 2.3 Our Contribution: A Separation Oracle with Restricted Support

Our solution to 2 is based on a new ellipsoid algorithm for finding separating constraints with restricted support. Specifically, the main technical contribution of this paper is Algorithm 1 with the following guarantee: Given a feasibility problem of the form

$$\text{Find } y \in \mathbb{R}^n \text{ subject to } \Pi y = y_0, \; y \in \text{int}(K),$$

where $\Pi$ is a projection matrix, $y_0 \in \text{span}(\Pi) \subseteq \mathbb{R}^n$; along with separation oracle for convex body $K$; it either finds feasible $y$ or asserts that the problem is infeasible and outputs a separating hyperplane $c \in \text{span}(\Pi)$. This algorithm coupled with the recursive separation oracle meets both our correctness and running time requirements. In particular, the running time instead of being the trivial bound of $|N|^{O(1)}$ will be roughly $|S(\ell)|^{O(\ell)}$. Assuming the exponential-time hypothesis, the exponential dependence on the number of seed selection stages $\ell$ cannot be avoided (a sub-exponential dependence would lead to a $f(k)n^{o(k)}$ time algorithm to decide if an $n$-vertex graph has a $k$-clique).

**Remark 1.** Our algorithm can be thought as a weak separation oracle for eq. (1) at level $i$ given a weak separation oracle for level $i+1$. When each convex body $K_{S(2)}, \ldots, K_{S(\ell)}$ is guaranteed to be a polytope, such as LPs from the Sherali-Adams Hierarchy, it is known that one can obtain a strong separation oracle at level $i$ by using only using a strong separation oracle at level $i+1$ (see Corollary 6.5.13 in [GLS93]). However in the case of semi-definite programming, it is an open question [PK97] whether one can obtain a strong separation oracle from another strong separation oracle in polynomial time.

## 3 Preliminaries

For any positive integer $n$, let $[n] \triangleq \{1, 2, \ldots, n\}$. We will use $\emptyset$ to denote empty set. Given set $A$, let $2^A$ be its power set, i.e. set of all subsets. For any real $k$, we will use $A_k, A_{\leqslant k}$ and $A_{\geqslant k}$ to denote the set of all subsets of $A$ having size exactly $k$, at most $k$ and at least $k$ respectively. Observe that $2^A = A_{\geqslant 0}$, $\emptyset = A_0$. Finally note that $A_{\geqslant 1}$ is the set of non-empty subsets of $A$.

Given sets $A, B$ and a field $\mathbb{F}$ ($\mathbb{R}$ for reals, $\mathbb{Q}$ for rationals), we will use $\mathbb{F}^A$ and $\mathbb{F}^{A \times B}$ to denote vectors and matrices over $\mathbb{F}$ whose rows and columns are identified with elements of $A$ and $B$ respectively. For any function $f : A \to \mathbb{F}$ (resp. $g : A \times B \to \mathbb{F}$), we will use $[f(u)]_{u \in A}$ (resp. $[g(u,v)]_{(u,v) \in A \times B}$) to denote the vector (resp. matrix) whose value at row $u$ (resp. row $u$ and column $v$) is equal to $f(u)$ (resp. $g(u,v)$). Given vector $x \in \mathbb{F}^A$ and matrix $Y \in \mathbb{F}^{A \times B}$, for any subset $C$ and $D$ let $x_C \in \mathbb{F}^C$ and $Y_{C,D} \in \mathbb{F}^{C \times D}$ denote the minors of $x$, $Y$ on rows $A \cap C$ and columns $B \cap D$ with 0's everywhere else.

Finally we will use $\mathbf{S}^A \supset \mathbf{S}^A_+ \supset \mathbf{S}^A_{++}$ to denote the set of symmetric, positive semi-definite and positive definite matrices on rows and columns $A$.

## 3.1 Convex Geometry and Ellipsoid Method

The main crux of our algorithm relies on an ellipsoid solver method which can also return a certificate of infeasibility. Throughout this section, we assume the underlying space is $n$-dimensional whose coordinates are identified with $[n]$. So all our vectors and matrices (unless noted otherwise) will have $[n]$ as their rows.

**Notation 5** (Projection). *We will use $\Pi \in \mathbf{S}_+^{[n]}$ to denote a projection matrix representing some linear subspace $\mathrm{span}(\Pi) \subseteq \mathbb{R}^{[n]}$ and $\Pi^\perp$ to denote the projection matrix onto null space of $\Pi$, i.e. $\Pi^\perp = \mathrm{identity} - \Pi$.*

*Given vector $y_0 \in \mathbb{R}^{[n]}$, we will use $y_0 \in \Pi$ if $y_0$ is in the span of $\Pi$, i.e. $\Pi y_0 = y_0$ and we will use $\Pi^{-1}(y_0)$ to denote the following set of vectors:*

$$\Pi^{-1}(y_0) \triangleq \left\{ y \in \mathbb{R}^{[n]} \middle| \Pi^\perp(y - y_0) = 0 \right\}.$$

**Notation 6** (Balls). *Given a set $K \subseteq \mathbb{R}^{[n]}$ and non-negative real $\varepsilon \geqslant 0$, we define $\mathbb{B}(K, \pm\varepsilon)$ in the following way.*

$$\mathbb{B}(K, \varepsilon) \triangleq \left\{ x \in \mathbb{R}^{[n]} | \exists y \in K \ s.t. \ \|y - x\|_2 \leqslant \varepsilon \right\}.$$

$$\mathbb{B}(K, -\varepsilon) \triangleq K \setminus \mathbb{B}(\mathbb{R}^{[n]} \setminus K, \varepsilon).$$

*Observe that for $y \in \mathbb{R}^{[n]}$, $\mathbb{B}(y, \varepsilon)$ is the n-dimensional sphere with origin $y$, with $\mathbb{B}(K, \varepsilon)$ being Minkowski addition of sphere of radius $\varepsilon$ to $K$ and $\mathbb{B}(K, -\varepsilon)$ being Minkowski subtraction of sphere of radius $\varepsilon$ from $K$.*

**Observation 7** (See [GLS93]). *For any convex body $K \subseteq \mathbb{R}^{[n]}$ and non-negative reals $\varepsilon, \varepsilon_1, \varepsilon_2$, the following hold:*

1. $\mathbb{B}(\mathbb{B}(K, \varepsilon), -\varepsilon) = K$, $\mathbb{B}(\mathbb{B}(K, -\varepsilon), \varepsilon) \subseteq K$.
2. $\mathbb{B}(\mathbb{B}(K, \varepsilon_1), \varepsilon_2) = \mathbb{B}(K, \varepsilon_1 + \varepsilon_2)$.
3. $\mathbb{B}(\mathbb{B}(K, -\varepsilon_1), -\varepsilon_2) = \mathbb{B}(K, -\varepsilon_1 - \varepsilon_2)$.

**Notation 8** (Volumes). *Given $K \subseteq \mathbb{R}^{[n]}$, we will use $\mathrm{vol}_d(K)$ to denote d-dimensional volume of $K$, provided it exists. Furthermore for any non-negative real $\varepsilon \geqslant 0$, let $\mathrm{vol}_d(\varepsilon)$ be the volume of d-dimensional ball of radius $\varepsilon$. We will use $\mathrm{vol}_d^{-1}(K)$ to denote the radius of a d-dimensional sphere whose volume is equal to $\mathrm{vol}_d(K)$ so that*

$$\mathrm{vol}_d(K) = \mathrm{vol}_d(\mathrm{vol}_d^{-1}(K)).$$

**Notation 9** (Polytope). *Given matrix $A \in \mathbb{R}^{m \times n}$ and vector $b \in \mathbb{R}^m$, let*

$$\mathrm{poly}(A, b) \triangleq \left\{ x \in \mathbb{R}^{[n]} | Ax \leqslant b \right\}.$$

**Lemma 10** (See Lemma 3.2.35 in [GLS93]). *Given a polytope $P = \mathrm{poly}(A, b)$, for any positive real $\varepsilon > 0$,*

$$\mathbb{B}(P, -\varepsilon) = \mathrm{poly}(A, b - \varepsilon\sqrt{\mathrm{diag}(A^T A)}).$$

*Here $\sqrt{\mathrm{diag}(A^T A)}$ denotes the vector whose $i^{th}$ coordinate is equal to Euclidean norm of $i^{th}$ row of $A$.*

**Definition 11** (Separation Oracle). *Given a convex body $K \subseteq \mathbb{R}^{[n]}$, $\mathsf{SEP}_\delta(y)$ is a separation oracle for $K$ if the following holds. On inputs a rational vector $y \in \mathbb{Q}^{[n]}$ and rational number $\delta > 0$, $\mathsf{SEP}_\delta(y)$ asserts feasible if $y \in K$. Otherwise, if $y \notin K$, it returns $c$ such that $\|c\|_\infty = 1$ and*

$$\forall x \in K : \langle c, x \rangle \leq \langle c, y \rangle + \delta.$$

*We will use $T(\mathsf{SEP}_\delta)$ to denote the worst case running time of $\mathsf{SEP}_\delta$.*

**Theorem 12** (Central-Cut Ellipsoid Method [GLS93]). *There exists an algorithm, called the central-cut ellipsoid method,*

$$\mathsf{CCUT\text{-}E}(\mathsf{SEP}_\delta, \Pi, y_0, \varepsilon_0)$$

*that solves the following problem. Given a projection matrix $\Pi \in \mathbf{S}_+^{[n]}$ of rank $m$, vector $y^0 \in \Pi$, a convex body $K \subseteq [-\Delta, \Delta]^{[n]}$ for some positive $\Delta$ with $\mathsf{SEP}_\delta$ (see Definition 11) and rational number $\varepsilon_0 > 0$, it runs in time*

$$|\log \Delta| N \left[\mathrm{poly}(n) + T(\mathsf{SEP}_{2^{-N}})\right] \text{ where } N \leq 6(n-m)(|\log \varepsilon_0| + (n-m));$$

*after which it outputs:*

1. *Either a vector $a \in \mathbb{Q}^{[n]}$ such that $a \in K \cap \Pi^{-1}(y^0)$;*
2. *Or a polytope of the form $P = \mathrm{poly}(C, d)$, where $C \in \mathbb{Q}^{[N] \times [n]}, d \in \mathbb{Q}^{[N]}$ with $K \subseteq P$ and $\mathrm{vol}_{n-m}\left(P \cap \Pi^{-1}(y^0)\right) < \varepsilon_0$.*

*Proof.* Such algorithm can be obtained by trivial modifications to the central-cut ellipsoid algorithm [GLS93], which we outline here. Handling the constraint $\Pi a = y^0$ can be done by projecting the covariance matrix of ellipsoid onto $\Pi^{-1}(y^0)$. At $k^{th}$ iteration, for all $k$, we add hyperplanes returned by $\mathsf{SEP}_\delta$ to $P$.

Algorithm terminates with a feasible $a \in \mathbb{Q}^{[n]}$ with $\Pi a = y^0$ only if $\mathsf{SEP}_\delta(a)$ asserts feasible for some $\delta \leq \varepsilon_0$, in which case $a \in K$. Otherwise, when the maximum number of iterations is reached we simply return. $\square$

## 4 Finding Separating Hyperplanes on a Subspace

We now describe our main technical contribution: An ellipsoid algorithm which can output a certificate of infeasibility **on a restricted subspace** using only the separation oracle $\mathsf{SEP}_\delta$ as in Definition 11. The procedure uses the central-cut ellipsoid method [GLS93] as a sub-routine. The main technical ingredient of our algorithm is Theorem 17, which is stated and proven in Section 4.1: It allows us to express this as another convex programming problem in terms of the "history" of constraints returned by separation oracle. Finally in Section 4.2 we present our ellipsoid algorithm, bound its running time and prove its correctness.

### 4.1 An Equivalent Convex Problem

We first state some useful propositions. Recall our goal: Given convex body $K$, a subspace $\Pi$ and $y_0 \in \Pi$, we have a polytope $P$ separating various points $\{y\} \subset \Pi^{-1}(y_0)$ from $K$. We want to compute a separating hyperplane on $\Pi$. Our approach is formulated in Lemma 16, see also Figure 1. We first show that points interior in $K$ have far off projections from $\Pi y_0$.

**Lemma 13.** *Given convex body $K \subseteq \mathbb{R}^{[n]}$, a projection matrix $\Pi \in \mathbf{S}_+^{[n]}$ with $\operatorname{rank}(\Pi) = m$, vector $y_0 \in \mathbb{R}^{[n]}$ and positive real $\delta > \operatorname{vol}_{n-m}^{-1}\left(\Pi^{-1}(y_0) \cap K\right)$,*

$$\text{for all } y \in \mathbb{B}(K, -2\delta), \ \|\Pi(y - y_0)\| \geqslant \delta.$$

We can restate our goal as the following: Given convex body $K$, a subspace $\Pi$ and $y_0 \in \Pi$, we have a polytope $P$ separating various points $\{y\} \subset \Pi^{-1}(y_0)$ from $K$. We want to show that points interior in $K$ have far off projections from $\Pi y_0$. First, we show that points in the interior cannot project exactly onto $\Pi y_0$.

**Proposition 14.** *Given $K \subseteq \mathbb{R}^{[n]}$, a projection matrix $\Pi \in \mathbf{S}_+^{[n]}$ with $\operatorname{rank}(\Pi) = m$, and $y \in \mathbb{R}^{[n]}$ the following holds: For any $\delta > \operatorname{vol}_{n-m}^{-1}(\Pi^{-1}(y) \cap K)$,*

$$\Pi^{-1}(y) \cap \mathbb{B}(K, -\delta) = \emptyset.$$

*Proof.* If $\exists x \in \Pi^{-1}(y) \cap \mathbb{B}(K, -\delta)$, then Observation 7 implies $\mathbb{B}(x, \delta) \subseteq K$. In particular,

$$\mathbb{B}(x, \delta) \cap \Pi^{-1}(y) \subseteq \Pi^{-1}(y) \cap K \implies \operatorname{vol}_m(\Pi^{-1}(y) \cap K) \geqslant \operatorname{vol}_m(\Pi^{-1}(y) \cap \mathbb{B}(x, \delta)).$$

Finally since $x \in \Pi^{-1}(y)$, $\mathbb{B}(x, \delta) \cap \Pi^{-1}(y)$ is an $m$-dimensional ball of radius $\delta$, whose volume is $\operatorname{vol}_m(\delta) > \operatorname{vol}_m(\Pi^{-1}(y) \cap K)$. Hence

$$\operatorname{vol}_m(\Pi^{-1}(y) \cap K) \geqslant \operatorname{vol}_m\left(\mathbb{B}(x, \delta) \cap \Pi^{-1}(y)\right) > \operatorname{vol}_m(\Pi^{-1}(y) \cap K),$$

which is a contradiction. $\square$

Lemma 13 is simply a quantitative version of the above, showing that points further interior in $K$ have far off projections from $\Pi y_0$.

**Lemma 15** (Lemma 13 restated). *Given convex body $K \subseteq \mathbb{R}^{[n]}$, a projection matrix $\Pi \in \mathbf{S}_+^{[n]}$ with $\operatorname{rank}(\Pi) = m$, vector $y_0 \in \mathbb{R}^{[n]}$ and positive real $\delta > \operatorname{vol}_{n-m}^{-1}\left(\Pi^{-1}(y_0) \cap K\right)$,*

$$\text{for all } y \in \mathbb{B}(K, -2\delta), \ \|\Pi(y - y_0)\| \geqslant \delta.$$

*Proof.* For the sake of contradiction, assume there exists $y \in \mathbb{B}(K, -2\delta)$ such that $\|\Pi(y - y_0)\| < \delta$. Consequently $\Pi\mathbb{B}(y, \delta)$, which is a sphere of radius $\delta$ on span of $\Pi$, contains $\Pi y_0$. In other words,

$$\emptyset \neq \Pi^{-1}(y_0) \cap \mathbb{B}(y, \delta). \tag{3}$$

Since $y \in \mathbb{B}(K, -2\delta)$, by convexity of $K$, we can repeatedly apply Observation 7 to show that

$$y \in \mathbb{B}(K, -2\delta) = \mathbb{B}(\mathbb{B}(K, -\delta), -\delta),$$
$$\mathbb{B}(y, \delta) \subseteq \mathbb{B}(\mathbb{B}(\mathbb{B}(K, -\delta), -\delta), \delta) \subseteq \mathbb{B}(K, -\delta).$$

Substituting this into eq. (3), we have $\emptyset \neq \Pi^{-1}(y_0) \cap \mathbb{B}(y, \delta) \subseteq \Pi^{-1}(y_0) \cap \mathbb{B}(K, -\delta)$ which contradicts Proposition 14 for our choice of $\delta$. $\square$

Having shown that there is a $\delta$-neighborhood of $\Pi y_0$ disjoint from interior of $K$ whenever their intersection has small volume, we can immediately use Minkowski's Separating Hyperplane Theorem to infer the existence of such hyperplane. In fact, any hyperplane perpendicular to the line from $y_0$ to the closest point in $K$ has this property. We formalize this below.

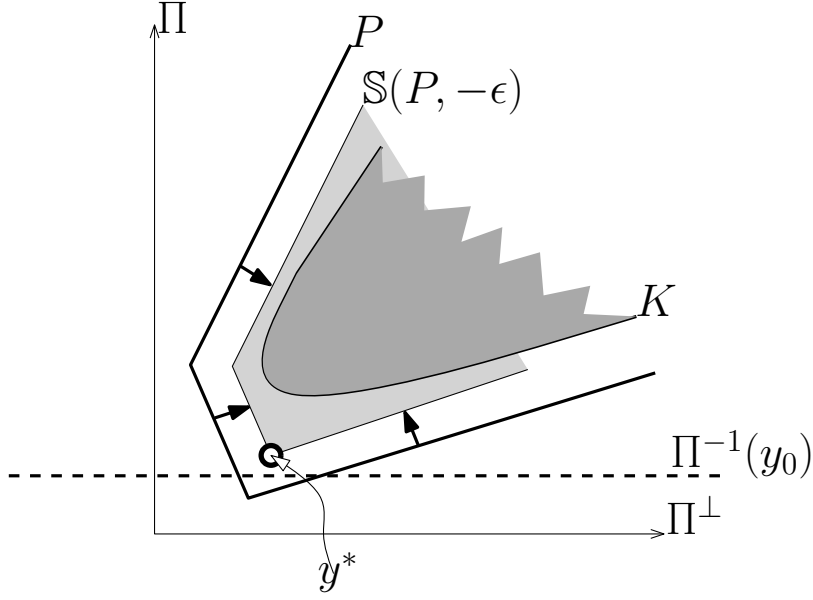

Figure 1: We want to find a hyperplane parallel to $\Pi^\perp$ separating $\Pi^{-1}(y_0)$ and $K$, using only the inequalities returned by separation oracle, polytope $P$. The optimal solution of Lemma 16 is given by $y^*$ with corresponding hyperplane $\Pi^{-1}(y^*)$.

**Lemma 16.** *Given convex body $K \subseteq \mathbb{R}^{[n]}$, a projection matrix $\Pi \in \mathbf{S}_+^{[n]}$ with $\mathrm{rank}(\Pi) = m$, vector $y_0 \in \mathbb{R}^{[n]}$ and positive real $\delta > \mathrm{vol}_{n-m}^{-1}\left(\Pi^{-1}(y_0) \cap K\right)$, the hyperplane perpendicular to the projection of direction from $y$ to closest point in the interior of $\Pi K$ separates $y_0$ and interior of $\Pi K$:*
*Formally any optimal solution $y^*$ to the following eq. (4):*

$$\begin{array}{ll} \text{Minimize} & \|\Pi(y - y_0)\|^2 \\ \text{subject to} & y \in \mathbb{B}(K, -2\delta), \end{array} \qquad (4)$$

*satisfies eq. (5):*

$$\min_{x \in \mathbb{B}(K, -2\delta)} \langle \Pi(y^* - y_0), x - y_0 \rangle \geqslant \|\Pi(y^* - y_0)\|^2. \qquad (5)$$

*Proof.* By contradiction. Assume for optimal solution $y^*$, there exists $x \in \mathbb{B}(K, -2\delta)$ such that

$$\langle \Pi(y^* - y_0), x - y_0 \rangle < \|\Pi(y^* - y_0)\|^2 = \langle \Pi(y^* - y_0), \Pi(y^* - y_0) \rangle = \langle \Pi(y^* - y_0), y^* - y_0 \rangle \,.$$

Therefore

$$\langle \Pi(y^* - y_0), x - y^* \rangle < 0 \qquad (6)$$

For some $\theta \in (0, 1]$ to be chosen later, consider $y(\theta) \leftarrow (1 - \theta) \cdot y^* + \theta \cdot x$, which is always feasible for eq. (4) as $x \in \mathbb{B}(K, -2\delta)$ and $\mathbb{B}(K, -2\delta)$ is convex. Then:

$$\begin{aligned} \frac{1}{2} \left. \frac{\partial \|\Pi(y(\theta) - y_0)\|^2}{\partial \theta} \right|_{\theta \to 0^+} &= \left\langle \Pi(y(\theta) - y_0), \Pi \frac{\partial(y(\theta) - y_0)}{\partial \theta} \right\rangle \bigg|_{\theta \to 0^+} \\ &= \left\langle \Pi(y(\theta) - y_0), \frac{\partial(y(\theta) - y_0)}{\partial \theta} \right\rangle \bigg|_{\theta \to 0^+} \\ &= \langle \Pi(y(\theta) - y_0), x - y^* \rangle|_{\theta \to 0^+} = \langle \Pi(y(0) - y_0), x - y^* \rangle \\ &= \langle \Pi(y^* - y_0), x - y^* \rangle < 0 \end{aligned}$$

where we used eq. (6) at the last step. We arrive at a contradiction by noting that above inequality implies existence of $\theta^* \in (0, 1]$ such that

$$\|\Pi(y(\theta^*) - y_0)\|^2 < \|\Pi(y^* - y_0)\|^2, \ y(\theta^*) \in \mathbb{B}(K, -2\delta). \qquad \square$$

Given Lemma 16, we can choose our separating hyperplane $c$ as $c = -\frac{\Pi(y^* - y_0)}{\|\Pi(y^* - y_0)\|_\infty}$. But this does not quite work for two reasons:

1. Hyperplane $c$ only separates "strict interior" of $K$ as it is, whereas we need to separate $K$ itself.
2. Depending on $K$, it might not be possible to represent optimal $c$ using polynomially many bits, thus we need to account for near optimal solutions.

We now show how to overcome these problems.

**Theorem 17.** *Given convex body $K \subseteq [-\Delta, \Delta]^n$ for some $\Delta > 0$, a projection matrix $\Pi \in \mathbf{S}_+^{[n]}$ with $\mathrm{rank}(\Pi) = m$, a vector $y_0 \in [-\Delta, \Delta]^n$, for any positive real $\delta > 0$ with $\delta > \mathrm{vol}_{n-m}^{-1}\left(\Pi^{-1}(y_0) \cap K\right)$, the following holds: If $y'$ is an $\frac{\delta^2}{2\Delta\sqrt{m}}$- approximate solution to eq. (7)*

$$\begin{aligned} \text{Minimize} & \quad \|\Pi(y - y_0)\|^2 \\ \text{subject to} & \quad y \in \mathbb{B}\left(K, -\left(2 + \frac{\delta}{2\sqrt{m}\Delta}\right) \cdot \delta\right) \end{aligned} \qquad (7)$$

*then $\Pi(y' - y_0) \neq 0$ and for $c$ being*

$$c \triangleq -\frac{\Pi(y' - y_0)}{\|\Pi(y' - y_0)\|_\infty} \implies \forall x \in K : \langle c, x \rangle \leqslant \langle c, y_0 \rangle + 2\delta\sqrt{m}. \qquad (8)$$

*Proof.* Before we begin, we set $\varepsilon \triangleq \frac{\delta}{2\sqrt{m}\Delta}$. Let $y^*$ be an optimal solution of eq. (7) with $\|y^* - y'\| \leqslant \varepsilon\delta$. Since $y^* \in \mathbb{B}(K, -(2+\varepsilon)\delta)$, we have $\mathbb{B}(K, -2\delta) \supseteq \mathbb{B}(y^*, \varepsilon\delta) \ni y'$. By Lemma 13, this implies $\|\Pi(y' - y_0)\| \geqslant \delta$, proving $\Pi(y' - y_0) \neq 0$. For any $x \in K$, we can decompose $x$ as $x = x' + z$ for some $x' \in \mathbb{B}(K, -2\delta)$ and $\|z\|_2 \leqslant 2\delta$. Then:

$$\begin{aligned} \langle \Pi(y' - y_0), x - y_0 \rangle =& \langle \Pi(y' - y_0), x' - y_0 \rangle + \langle \Pi(y' - y_0), z \rangle \\ \geqslant & \langle \Pi(y' - y^*), x' - y_0 \rangle + \langle \Pi(y^* - y_0), x' - y_0 \rangle - \|z\| \cdot \|\Pi(y' - y_0)\| \\ \geqslant & - \underbrace{\|\Pi(y' - y^*)\|}_{\leqslant \varepsilon\delta} \underbrace{\|\Pi(x' - y_0)\|}_{\leqslant 2\Delta\sqrt{m}} + \underbrace{\delta^2}_{\text{by Lemma 16}} - 2\delta\|\Pi(y' - y_0)\| \\ \geqslant & - 2\delta\|\Pi(y' - y_0)\| \text{ (by the choice of } \varepsilon = \frac{\delta}{2\Delta\sqrt{m}}) \\ \geqslant & - 2\delta\sqrt{m}\|\Pi(y' - y_0)\|_\infty. \end{aligned}$$

Since $c = -\frac{\Pi(y' - y_0)}{\|\Pi(y' - y_0)\|_\infty}$, we have $\langle c, x \rangle \leqslant \langle c, y_0 \rangle + 2\delta\sqrt{m}$ for any $x \in K$. $\qquad \square$

### 4.2 Ellipsoid Algorithm with Certificate of Infeasibility

Our solver is given in Algorithm 1.

**Algorithm 1** CERTIFY-E($\mathsf{SEP}_\delta, \Pi, y_0, \varepsilon_0$): Ellipsoid method with certificate of infeasibility.

**Input.** • Convex body $K \subseteq [0,1]^n$ and separation oracle $\mathsf{SEP}_\delta$ as in Definition 11,
• Projection matrix $\Pi \in \mathbf{S}_+^{[n]}$ with rank($\Pi$) = $m$, $y_0 \in \mathbb{R}^{[n]}$ and positive real $0 < \varepsilon_0 < 1$.

**Output.** • Either a vector $y \in \mathbb{Q}^n$ st $y \in K \cap \Pi^{-1}(y_0)$.
• Or $c \in \Pi$ with $\|c\|_\infty = 1$ and $\forall x \in K : \langle c, x \rangle \leq \varepsilon_0 + \langle c, y_0 \rangle$.

**Procedure.** 1. Run CCUT-E($\mathsf{SEP}_\delta, \Pi, y_0, \varepsilon$) where $\varepsilon \leftarrow \mathrm{vol}_{n-m}\left(\frac{\varepsilon_0}{2\sqrt{m}}\right)$.

2. If it returns $y \in K \cap \Pi^{-1}(y_0)$, then return $y$.

3. Else let $P = \mathrm{poly}(C, d)$ be the polytope it returns. Set $\delta \leftarrow \frac{\varepsilon_0}{2\sqrt{m}}$. $\varepsilon' \leftarrow \frac{\delta}{2\Delta\sqrt{m}}$ and

4. Solve eq. (9) using regular ellipsoid method to find an $\varepsilon'\delta$-approximate solution, $y^* \in \mathbb{Q}^n$:

$$\text{Minimize } \|\Pi(y - y_0)\|^2 \text{ subject to } Cy \leq d - (2 + \varepsilon')\delta\sqrt{\mathrm{diag}(C^T C)}. \tag{9}$$

5. Return $c \leftarrow -\frac{\Pi(y^* - y_0)}{\|\Pi(y^* - y_0)\|_\infty}$.

---

The proof of the following theorem follows by combining various ingredients we have so far, especially Theorems 12 and 17.

**Theorem 18** (Main technical tool). *Algorithm 1 runs in time $N \cdot T(\mathsf{SEP}_{2^{-N}}) + \mathrm{poly}(n)\log^2 \frac{1}{\varepsilon_0}$ where $N = O\left((\# \text{ of free variables})^2 \log \frac{\# \text{ of fixed variables}}{\varepsilon_0}\right) = O\left((n-m)^2 \log \frac{m}{\varepsilon_0}\right)$, and provides the following guarantee: If $\mathrm{vol}_{n-m}^{-1}(K \cap \Pi^{-1}(y_0)) > \frac{\varepsilon_0}{2\sqrt{m}}$ then it outputs $y \in K \cap \Pi^{-1}(y_0)$.*

*Proof of Running Time.* By Theorem 12, step 1 finishes in time $N(\mathrm{poly}(n) + T(\mathsf{SEP}_{2^{-N}}))$ with $N = O((n-m)^2 + (n-m)\log 1/\varepsilon)$, where $\log 1/\varepsilon = \log 1/\mathrm{vol}_{n-m}(\varepsilon_0/2\sqrt{m}) \leq O\left((n-m)\log \frac{m}{\varepsilon_0}\right)$ so $N = O\left((n-m)^2 \log \frac{m}{\varepsilon_0}\right)$.

For step 4, we can implement a simple separation oracle which runs in time $O(Nn)$. The regular ellipsoid method requires $O(n^2 + n\log(1/\varepsilon'\delta))$ iterations to reach an accuracy of $\varepsilon'\delta$. Each iteration takes time $\mathrm{poly}(n)$ in addition to separation oracle, therefore the total running time of fourth step is bounded by $N\mathrm{poly}(n)\log(1/\varepsilon_0) = \mathrm{poly}(n)\log^2 \frac{1}{\varepsilon_0}$. Hence the claim follows. □

*Proof of Correctness.* There are two cases. If algorithm outputs $y$ at step 2, by Theorem 12 $y \in \mathbb{Q}^n \cap \Pi^{-1}(y_0) \cap K$.

Now consider the other case. Then step 1 will output a polytope $P = \mathrm{poly}(C, d)$ such that $K \subseteq P$, whose volume is bounded by

$$\mathrm{vol}_{n-m}(P \cap \Pi^{-1}(y_0)) < \mathrm{vol}_{n-m}\left(\frac{\varepsilon_0}{2\sqrt{m}}\right).$$

The set of feasible solutions for eq. (9) is $\mathbb{B}(P, -(2+\varepsilon')\delta)$ by Lemma 10. Since $\mathrm{vol}_{n-m}^{-1}(P \cap \Pi^{-1}(y_0)) < \delta := \frac{\varepsilon_0}{2\sqrt{m}}$, we can apply Theorem 17 and conclude that $c$ as constructed in step 5 will have the following properties:

- $c \in \Pi$,

- $\|c\|_\infty = 1$,
- For all $x \in P$, $\langle c, x \rangle \leq \langle c, y_0 \rangle + 2\sqrt{m}\delta$. To see this, note $K \subseteq P$, for all $x \in K$ means:

$$\langle c, x \rangle \leq \max_{x \in K}\langle c, x \rangle \leq \max_{x' \in P}\langle c, x' \rangle \leq \langle c, y_0 \rangle + 2\sqrt{m}\delta = \langle c, y_0 \rangle + \varepsilon_0.$$

This finishes the proof of correctness. $\square$

## 5 Faster Solver for Local Rounding Algorithms

We return back to our motivating example. Assume we have $n$ variables, and we want to find a discrete labeling $\widetilde{x} \in L^{[n]}$ of those from a set of labels $L$, under various constraints and objective. Suppose we "lifted" this problem into a higher dimension $\mathbb{R}^N$ where $|N| \gg n$, and obtained a family of increasingly tight convex relaxations defined over various subspaces of $\mathbb{R}^N$.

Formally, we have a set of subspaces $\{\Pi_S\}_{S \subseteq [N]}$, represented by their projection matrices and associated with subsets of $[N]$, and with each subspace $\Pi_S$, we have an associated convex body, $K_S \subseteq \Pi_S[0,1]^N$ with such that

$$\Pi_S \subseteq \Pi_T \implies \Pi_S K_T \subseteq K_S.$$

We are given functions FEASIBLE, ROUND and SEED, along with positive integers $n, s$ such that:

- FEASIBLE$_S$ : $\Pi_S \mathbb{Q}^N \to \{\text{feasible}, \Pi_S \mathbb{Q}^N\}$. On input $S \subseteq N, y \in \Pi_S Q^N$, it asserts feasible if $y \in K_S$ or returns $c \in \Pi_S \mathbb{Q}^N : \|c\|_\infty = 1$ such that[4] $\forall x \in K_S : \langle c, x \rangle < \langle c, y \rangle$ in time poly(rank($\Pi_S$)).

- SEED$_S$ : $K_S \to 2^N$. Given $S \subseteq [N]$ and $y \in \Pi_S \mathbb{Q}^N$, it returns subset $S' \supseteq S$ such that $\frac{\text{rank}(\Pi_{S'})}{\text{rank}(\Pi_S)} \leq s$ when $S \neq \emptyset$, and rank$(\Pi_{S'}) \leq n$ when $S = \emptyset$. Its worst case running time is bounded by poly(rank($\Pi_S$)) (or poly($n$) in the case of $S = \emptyset$).

- ROUND$_S$ : $K_S \to L^{[n]}$. On inputs $S \subseteq N$ and $y \in K_S$, returns an approximation to the original problem in time poly(rank($\Pi_S$)).

We now describe our main solver. Note that once the algorithm outputs $y^* \in K_{S(\ell)}$, the final output labeling will simply be ROUND$_{S(\ell)}(y^*)$. The proof of the following theorem is easy given the ingredients so far.

**Theorem 19.** *Algorithm 2 runs in time $\left[s^\ell n \log(1/\varepsilon_0)\right]^{O(\ell)}$ (compare this with the straightforward algorithm which runs in time $N^{O(1)} \log(1/\varepsilon_0)$) and achieves the following guarantee: Provided that $\text{vol } K > \varepsilon_0$, it outputs $y^* \in K_{S(\ell)}$ and $S(0), \ldots, S(\ell)$ st for all $i$:*

$$\Pi_{S(i)} y^* \in K_{S(i)}, \tag{10}$$
$$S(i+1) = \textsf{SEED}_{S(i)}(y^*). \tag{11}$$

*Otherwise it asserts $\text{vol } K \leq \varepsilon_0$.*

*Furthermore there is no algorithm which runs in time $n^{o(\ell)}$ assuming Exponential Time Hypothesis.*

---
[4] We can handle FEASIBLE$_S$ that only returns a weak separation oracle, but since in our application to SDPs we have access to a strong separation oracle, we assume this for simplicity.

**Algorithm 2** Fast Solver (to fool local rounding algorithm)

**Input.** • Maximum number of iterations $\ell$ and positive real $\varepsilon_0 > 0$,
• $n$, $r$, $(K_S)_{S \subseteq [N]}$ with separation oracle FEASIBLE, SEED, $\Pi_\emptyset$ and $y(0) \in K_\emptyset \mathbb{Q}^N$ all as described in Section 5.

**Output.** • Either asserts $\text{vol } K \leqslant \varepsilon_0$,
• Or outputs $y^* \in K_{S(\ell)}$ and $S(0), \ldots, S(\ell)$ st for all $i$: 1. $\Pi_{S(i)} y^* \in K_{S(i)}$; 2. $S(i+1) = \text{SEED}_{S(i)}(y^*)$.

**Procedure.** 1. Initialize global variables $S(1), \ldots, S(\ell)$ representing seed sets and global sparse vector $y^* \in \mathbb{Q}^{[N]}$ representing the final solution (it will be in span of $\Pi_{S(\ell)}$).

2. Set $S(0) \leftarrow \{\emptyset\}$.

3. Run CCUT-E(SEP$_{S(0),\delta}, 0, 0, \varepsilon_0$) (see Theorem 12) where SEP is given in Algorithm 3.

4. If it asserts feasible, output $S(0), \ldots, S(\ell)$ and $y^*$.

5. Else assert $\text{vol } K \leqslant \varepsilon_0$.

---

We prove correctness, running time claim, and ETH hardness in turn.

*Proof of Correctness.* First we assume correctness of Algorithm 3 and prove correctness of Algorithm 2. There are only two cases:

1. Algorithm 2 returns $y^* \in K_{S(\ell)}$ only if CCUT-E$_{S(1)}$ returns a feasible solution. For such $y^*$, by Theorem 12, SEP$_{S(1),\varepsilon_0}(y^*)$ asserts feasible. By correctness of SEP$_{S(1),\varepsilon_0}(y^*)$, $y^*$ satisfies all claims.

2. Else algorithm asserts $\text{vol}_{[N]} K \leqslant \varepsilon_0$ which means CCUT-E asserted $\varepsilon_0 \geqslant \text{vol}_{S(0)} \mathbb{B}(K_{S(1)}, \varepsilon_0) \geqslant \text{vol}_{S(0)} K_{S(0)}$. We know that $\Pi_{S(0)} K \subseteq K_{S(0)}$ and $K \subseteq [0,1]^{[N]}$ therefore $\text{vol}_{[N]}(K) \leqslant \text{vol } \Pi_{S(0)} K \leqslant \text{vol}_{S(0)} K_{S(0)} \leqslant \varepsilon_0$.

Now we will prove correctness of Algorithm 3 inductively starting from $i = \ell$. For each $i$, in order to prove inductive step, we need to consider in which one of the following steps SEP$_{S(i),\varepsilon_0}$ returned:

1. Step 1: Follows from definition of FEASIBLE.

2. Steps 2 and 6: By construction of $S(j)$'s and inductive hypothesis, $S(j+1) = \text{SEED}_{S(j)}(y^*)$ holds for all $j \geqslant i$.

   We will prove that FEASIBLE$_{S(j)}(y^*)$ asserts feasible for all $j \geqslant i$. For $j > i$, this immediately follows from inductive hypothesis. For $j = i$, at Step 1 FEASIBLE$_{S(i)}(y^*)$ asserted feasible. Thus $\Pi_{S(j)} y^* \in K_{S(j)}$ for all $j$.

3. Step 5: It returns $c$, only if CERTIFY-E at step 4 outputs $c$, correctness of which follows from Theorem 18. □

*Proof of Running Time.* If we let $n_i \leftarrow ns^{i-1}$ for $i \geqslant 1$, we can see that $\text{rank}(\Pi_{S_i}) \leqslant n_i$ at $i^{th}$ iteration. Hence
$$T(\text{FEASIBLE}_{S(i)}) + T(\text{SEED}_{S(i)}) = (sn_i)^{O(1)} = n_{i+1}^{O(1)}.$$

**Algorithm 3** $\mathsf{SEP}_{S(i),\varepsilon_0}(y)$:Separation Oracle.

**Input.** • Positive real $\varepsilon_0 > 0$, current iteration $i$, current seeds $S(i)$, vector $y \in \mathbb{Q}^{S(i)}$.

**Output.** • Either asserts feasible, and sets values of global variables $S(i+1), \ldots, S(\ell)$ along with $y^*$ so that:

1. $\Pi_{S(j)} y^* \in K_{S(j)}$ for all $j : i \leqslant j \leqslant \ell$,
2. $S(j+1) = \mathsf{SEED}_{S(j)}(y^*)$ for all $j : i \leqslant j \leqslant \ell - 1$.

• Or returns $c \in \Pi_{S(i)}$ with $\|c\|_\infty = 1$ such that $\forall x \in K_{S(i)} : \langle c, x - y \rangle < \varepsilon_0$.

**Procedure.** 1. If $\mathsf{FEASIBLE}_{S(i)}(y)$ returns $c \in \Pi_{S(i)}$, return $c$.

2. Else if $i \geqslant \ell$, set $y^* \leftarrow y$. Assert feasible and return.
3. $S(i+1) \leftarrow \mathsf{SEED}_{S(i)}(y)$.
4. Run $\mathsf{CERTIFY\text{-}E}(\mathsf{SEP}_{\Pi_{S(i+1)},\delta}, \Pi_{S(i)}, y, \varepsilon_0)$ (see Algorithm 1).
5. If it returns $c$, return $c$.
6. Else assert feasible.

---

If we use $T_i$ to denote the maximum of $T(\mathsf{SEP}_{S(i),\varepsilon_0}(y))$ over all possible $S(i)$ and $y$'s, with $T_0 = T(\text{main})$; then $T_\ell = n_\ell^{O(1)} = (r^\ell n)^{O(1)}$ and for any $i < \ell$:

$$T_i \leqslant O(n_{i+1}^2 \log n_i/\varepsilon_0) T_{i+1} + n_{i+1}^{O(1)} = (n_{i+1})^{O(1)} \log(1/\varepsilon_0) T_{i+1} = s^{O(i)} n^{O(1)} \log 1/\varepsilon_0 \cdot T_{i+1}$$
$$T_0 = s^{O(\ell^2)} n^{O(\ell)} \log^\ell(1/\varepsilon_0). \qquad \square$$

*Proof of ETH Hardness.* Consider the $k$-clique problem, which cannot be solved in time $f(k)n^{o(k)}$ under the exponential-time hypothesis [LMS11]. Moreover it is easy to see that $k$ rounds of Lasserre Hierarchy relaxation is integral for this problem [Lau03], and we can find such a $k$-clique using a seed selection procedure with $\ell = k$ levels. $\qquad\square$

Next, we will review semidefinite programs from the Lasserre Hierarchy in Section 6 and finally in Section 7, we will show a sample of how some approximation algorithms using Lasserre Hierarchy relaxation fit into our framework.

## 6 Lasserre Hierarchy

In this section, we present a general derivation of Lasserre Hierarchy relaxation. Although our relaxation is only given for 0/1 programming problems, it can easily be adapted to work on any set of finite labels.

### 6.1 Preliminaries

For some positive integer $d$, let $Q \in \mathbb{R}^{[n]\leqslant d}$ (resp. $\mathcal{P} = \{P_1, \ldots, P_M\} \subset \mathbb{R}^{[n]\leqslant d}$) be a degree $\leqslant d$ multilinear polynomial (resp. subset of degree $\leqslant d$ multilinear polynomials) over $n$ variables representing objective function (resp. constraints). We want to find a binary labeling of $n$ variables,

$\widetilde{x} \in \{0,1\}^n$, which satisfies eq. (12):

$$\begin{array}{ll} \text{Minimize} & \sum_{S \in [n]_{\leq d}} Q_S \prod_{u \in S} \widetilde{x}_u \\ \text{subject to} & \sum_{S \in [n]_{\leq d}} P_S \prod_{u \in S} \widetilde{x}_u \geq 0 \quad \text{for all } P \in \mathcal{P}, \\ & \widetilde{x} \in \{0,1\}^n. \end{array} \qquad (12)$$

Observe that we can convert any constraint satisfaction problem on $\{0,1\}^{[n]}$ to this form easily. In this section, we will first express eq. (12) as an **equivalent** SDP problem, from which Lasserre relaxation can be obtained by enforcing positivity only on certain principal minors of this matrix. This exposition of Lasserre relaxation through constraints on certain minors will be convenient when we are presenting our partial SDP solver.

**Notation 20.** *Given positive integers $n, d$, for any vector $P \in \mathbb{R}^{[n]_{\leq d}}$ and $y \in \mathbb{R}^{2^{[n]}}$, we define $P * y \in \mathbb{R}^{2^{[n]}}$ as the following vector:*

$$(P * y)_S \triangleq \sum_{T \in [n]_{\leq d}} P_T y_{T \cup S}.$$

**Definition 21** (Multivariate Moment Matrix on $\mathcal{P}$ and $Q$)**.** *Given positive integer $n$ let $\mathbb{M}^n : \mathbb{R}^{2^{[n]}} \to \mathbb{R}^{2^{[n]} \times 2^{[n]}}$ be the following linear matrix function:*

$$\mathbb{M}^n(y) = [y_{A \cup B}]_{A, B \subseteq [n]}.$$

*For any $\mathcal{P} = \{P_1, \ldots, P_M\} \subseteq \mathbb{R}^{[n]_{\leq d}}$ representing constraints, $Q \in \mathbb{R}^{[n]_{\leq d}}$ and $q \in \mathbb{R}$ representing objective polynomial and a guess for its value, let $\mathbb{M}^{n, \mathcal{P}} : \mathbb{R}^{2^{[n]}} \to \mathbb{R}^{([m+2] \times 2^{[n]}) \times ([m+2] \times 2^{[n]})}$ be the following linear function on block diagonal matrices:*

$$\mathbb{M}^{n,\mathcal{P},Q,q}(y) \triangleq \begin{bmatrix} \mathbb{M}^n(y) & 0 & 0 & 0 & \cdots & 0 \\ 0 & qy_\emptyset - \langle Q, y \rangle & 0 & 0 & \cdots & 0 \\ 0 & 0 & \mathbb{M}^n(P_1 * y) & 0 & \cdots & 0 \\ 0 & 0 & 0 & \mathbb{M}^n(P_2 * y) & \cdots & 0 \\ \vdots & \vdots & \vdots & \vdots & \ddots & \vdots \\ 0 & 0 & 0 & 0 & \cdots & \mathbb{M}^n(P_M * y) \end{bmatrix}.$$

Having defined the moment matrix, we can state the following:

**Theorem 22** (See [Las02])**.** *eq. (12) has optimal value $\leq q$ if and only if there exists $y$ with $y \neq 0$ such that $\mathbb{M}^{n,\mathcal{P},Q,q}(y) \succeq 0$.*

**Definition 23** (Principal Minors of Multivariate Moment Matrices)**.** *Given $\mathcal{P} \subseteq \mathbb{R}^{[n]_{\leq d}}$, $Q \in \mathbb{R}^{[n]_{\leq d}}$ and $q \in \mathbb{R}$ as described, for any $T : T \subseteq 2^{[n]}, T \supseteq \mathrm{support}(Q)$ being a family of sets over $[n]$ containing the support of $Q$, we define $\mathbb{M}\big|_T^n(y) = \mathbb{M}\big|_T^{n,\mathcal{P},Q,q}(y)$ as the following principal minor of $\mathbb{M}^{n,\mathcal{P},Q,q}(y)$:*

$$\mathbb{M}\big|_T^n(y) \triangleq \begin{bmatrix} (\mathbb{M}^n(y))_{T,T} & 0 & 0 & \cdots & 0 \\ 0 & qy_\emptyset - \langle Q, y \rangle & 0 & \cdots & 0 \\ 0 & 0 & (\mathbb{M}^n(P_1 * y))_{T,T} & \cdots & 0 \\ \vdots & \vdots & \vdots & \ddots & \vdots \\ 0 & 0 & 0 & \cdots & (\mathbb{M}^n(P_M * y))_{T,T} \end{bmatrix}.$$

**Observation 24.** *Given a family of sets over $[n]$, $T \subseteq 2^{[n]}$, for any $P \in \mathbb{R}^{[n]_{\leqslant d}}$, the minor $(\mathbb{M}^n(P*y))_{T,T}$ is only a function of $y_{\mathrm{ex}(T,d)}$ where*

$$\mathrm{ex}(T, d) \triangleq \{A \cup B \cup C : A \in T, B \in T, C \in [n]_{\leqslant d}\}.$$

*Proof.* Consider $(\mathbb{M}^n(P*y))_{A,B}$ with $A, B \in U$. By Definition 21, this is equal to

$$(\mathbb{M}^n(P*y))_{A,B} = (P*y)_{A \cup B} = \sum_{C \in [n]_{\leqslant d}} P_C y_{A \cup B \cup C},$$

where $A \cup B \cup C \in \mathrm{ex}(U,d)$ by definition. The second part follows immediately from the definition of $\mathbb{M}\big|_T^n(y)$. □

Using Theorem 22 and observation 24, we can easily state and prove the following:

**Theorem 25** (Lasserre Relaxation [Las02]). *Given positive integers $n, r$ and $d$, polynomials $\mathcal{P} \subset \mathbb{R}^{[n]_{\leqslant d}}$ and $Q \in \mathbb{R}^{[n]_{\leqslant d}}$ with $q \in \mathbb{R}$, the following is $r^{th}$ round Lasserre Hierarchy relaxation of eq. (12):*

$$\text{Find } y \in \mathbb{R}^{[n]_{\leqslant 2r+d}} \text{ subject to } \mathbb{M}\Big|_{[n]_{\leqslant r}}^n (y) \succeq 0 \text{ and } y_\emptyset = 1. \tag{13}$$

Note that the straightforward SDP relaxation of eq. (12) corresponds to $d$ rounds of Lasserre hierarchy relaxation (and the "basic" SDP relaxation for the case of quadratic polynomials).

## 6.2 Separation Oracle for Lasserre Hierarchy

Given a binary labeling problem, we first cast $\tilde{r}$ rounds of Lasserre Hierarchy relaxation in our framework:

- The set of labels is $L = \{0, 1\}$.
- Lifted space $N$ is $\binom{[n]}{\leqslant r'}$, the subsets of $[n]$ of size at most $r'$,
- For any subset $S \subseteq [n]$, we define $\Pi_S$ as the projection matrix onto $\mathbb{R}^{\mathrm{ex}(S,2)}$ so that

$$[\Pi_S x]_T = \begin{cases} 1 & \text{if } T \in \mathrm{ex}(S, 2), \\ 0 & \text{else.} \end{cases}$$

The associated convex body, $K_S$, is defined as

$$K_S = \left\{ y \in \mathbb{R}^{\mathrm{ex}(S,2)} : y_\emptyset = 1, \ \mathbb{M}\Big|_{\mathrm{ex}(S,2)}^n (y) \succeq 0 \right\}.$$

Before stating the FEASIBLE procedure, we need the following well known result:

**Proposition 26.** *Given a symmetric matrix $A \in \mathbf{S}^B$, there exists an algorithm which asserts if $A \succeq 0$ or returns $x \in \mathbb{Q}^B$ such that $x^T A x < 0$ in time at most polynomial in size of $A$.*

Then our $\mathsf{FEASIBLE}_S(y)$ procedure is trivial: It asserts feasible if $\mathbb{M}\big|_{\mathrm{ex}(S,2)}^n (y) \succeq 0$. Else it returns $x \in \mathbb{Q}^{\mathrm{ex}(S,2)}$ for which $x^T \mathbb{M}\big|_{\mathrm{ex}(S,2)}^n (y) x < 0$.

# 7 Applications of faster solver to Lasserre hierarchy rounding algorithms

In this section, we finally apply our main algorithm as given in Algorithm 2 to various rounding algorithms for Lasserre Hierarchy relaxations as stated in the introduction. For all these problems, our separation oracle is the same procedure as described in Section 6.2. The running times we obtained as well as approximation factors and other guarantees are summarized in Figure 2. The last two columns list the value of $s$ (the factor by which $\text{rank}(\Pi_S)$ increases in each step of seed selection) and $\ell$ (the number of iterations of seed selection) used by the rounding algorithm in each case. The parameter $r$ refers to the index of the eigenvalue governing the approximation guarantee, and $\varepsilon$ is a positive parameter.

---

**Algorithm 4** $\text{SEED-QIP}_S(y)$: Seed selection procedure for approximation algorithms given in [GS11] for two-way partitioning problems.

**Input.** Subset $S \subseteq [n]$, and $y \in \mathbb{Q}^{\text{ex}(S,2)}$ provided $\mathbb{M}|_{\text{ex}(S,2)}^{[n]}(y) \succeq 0$ and $y_\emptyset = 1$, positive integer $r'$.

**Output.** $T$ with $|T| \leqslant r' \cdot |S|$.

**Procedure.** 1. Let $(x_T)_T$ be vectors corresponding to Cholesky factorization of $\mathbb{M}|_{\text{ex}(S,2)}^{[n]}(y)$. For each $S \in \text{ex}(S,2)$ and $f \in \{0,1\}^S$, set

$$x_S(f) \leftarrow \sum_{T \subseteq f^{-1}(0)} (-1)^{|T|} x_{f^{-1}(1) \cup T}. \tag{14}$$

2. Let $\Pi_S \leftarrow \sum_{f \in \{0,1\}^S} \frac{1}{\|x_S(f)\|^2} x_S(f) x_S(f)^T$ and $\Pi_S^\perp \leftarrow I - \Pi_S$.
3. Use volume sampling [DR10, GS12b] to choose $S'$: an $r'$-subset of vectors from $\left(\Pi_S^\perp x_u(1)\right)_{u \in [n]}$.
4. Return $S \cup S'$.

---

**Algorithm 5** $\text{SEED-COLOR}_S(y)$: Seed selection procedure for semi-coloring algorithm as given in [AG11] on graph $G$.

**Input.** Graph $G$ on nodes $[n]$, positive integer $r'$.

**Output.** $S \in [n]_{\leqslant r'}$.

**Procedure.** 1. Let $X_u \leftarrow \sum_{i=1}^{3} e_i \otimes x_\emptyset^\perp x_u(i)$ (same embedding as [AG11]).

2. Use volume sampling [DR10, GS12b] to choose $S$, an $r'$-subset of vectors from $(X_u)_{u \in [n]}$.

3. Return $S$.

---

The claimed running times follow from the $\approx s^{O(\ell^2)} n^{O(\ell)}$ runtime guaranteed by Theorem 19 for our solver (Algorithm 2). The rounding algorithm in each case runs within the same time. For problems marked with *, check the caption for required conditions.

The works [BRS11] and [AG11] use greedy seed selection, but these can be replaced by the above column selection procedure as well. Below, in Algorithms 4 to 6, we present seed selection procedures for binary graph partitioning algorithms from [GS11], sparsest cut algorithm from [GS12a], and semi-coloring algorithm from [AG11], respectively.

**Algorithm 6** SEED-SPARSEST-CUT$_S(y)$: Seed selection procedure for sparsest cut algorithm as given in [GS12a] on graphs $G$ and $H$.

**Input.** Graphs $G, H$ on nodes $[n]$ with edge weights given by $w_{u,v}^{G,H}$ respectively, positive integer $r'$.

**Output.** $S \in [n]_{\leqslant r'}$.

**Procedure.** 1. Let $X_{u,v} \leftarrow \sqrt{w_{u,v}^H}(x_u - x_v)$.
2. Use volume sampling [DR10, GS12b] to choose $S$, an $r'$-subset of vectors from $(X_{u,v})_{u,v \in [n]_2}$.
3. Return $S$.

| Problem Name | | Running Time | OPT | Rounding | $s$ | $\ell$ |
|---|---|---|---|---|---|---|
| Maximum Cut | [GS11] | $2^{O(r/\varepsilon^3)} n^{O(1/\varepsilon)}$ | $1 - \eta$ | $1 - \frac{1+\varepsilon}{\lambda_{n-r}} \eta$ | $2^{O(r/\varepsilon)}$ | $O(1/\varepsilon)$ |
| $k$-Unique Games | [GS11] | $k^{O(r/\varepsilon)} n^{O(1)}$ | $1 - \eta$ | $1 - \frac{2+\varepsilon}{\lambda_r} \eta$ | $k^{O(r/\varepsilon)}$ | $1$ |
| Minimum Bisection | [GS11] | $2^{O(r/\varepsilon^3)} n^{O(1/\varepsilon)}$ | $\eta$ | $\frac{1+\varepsilon}{\lambda_r} \eta - o(1)$ | $2^{O(r/\varepsilon)}$ | $O(1/\varepsilon)$ |
| Maximum Bisection | [GS11] | $2^{O(r/\varepsilon^3)} n^{O(1/\varepsilon)}$ | $1 - \eta$ | $1 - \frac{1+\varepsilon}{\lambda_{n-r}} \eta - o(1)$ | $2^{O(r/\varepsilon)}$ | $O(1/\varepsilon)$ |
| Sparsest Cut* | [GS12a] | $2^{O(r/\varepsilon)} n^{O(1)}$ | $\eta$ | $\frac{\eta}{\varepsilon}$ | $2^{O(r/\varepsilon)}$ | $1$ |
| Independent Set* | [GS11] | $2^{O(r)} n^{O(1)}$ | $\eta$ | $\Omega(\eta)$ | $2^{O(r)}$ | $O(1)$ |
| | [AG11] | $2^{O(r)} n^{O(1)}$ | $\eta$ | $\frac{\eta}{12}$ | $2^{O(r)}$ | $1$ |
| Maximum 2-CSPs* | [BRS11] | $2^{\text{poly}(k/\varepsilon) \cdot r} n^{\text{poly}(k/\varepsilon)}$ | $\eta$ | $\eta - \varepsilon$ | $k^{O(rk^2/\varepsilon^2)}$ | $O\left(k^2/\varepsilon^2\right)$ |

Figure 2: Running times and approximation guarantees for various Lasserre Hierarchy relaxation rounding algorithms using our faster solver. For sparsest cut, the spectral assumption is $\lambda_r \geqslant \eta/(1-\varepsilon)$. For independent set [GS11], the spectral assumption is $\lambda_{n-r} \leqslant 1 + O(1/\Delta)$ where $\Delta$ is the maximum degree. For independent set [AG11], the assumptions are that $G$ is 3-colorable and $\lambda_{n-r} \leqslant 17/16$. Finally for maximum 2-CSPs [BRS11], the assumption is that $\lambda_r \geqslant 1 - \left(\frac{\varepsilon}{2k}\right)^2$.

We conclude the paper by mentioning some notable Lasserre based approximation algorithms for which we are not able to get a runtime improvement:

- The algorithm for independent sets in 3-uniform hypergraphs [CS08] and the algorithm for directed Steiner tree [Rot11], which are adaptive with a large number of stages $\ell$ in the rounding procedure.

- The algorithm for minimum/maximum bisection in [RT12], which requires choosing the seed set at random *independently* from the final solution; whereas our solver, by nature, outputs a solution which depends on the seed set.

## Acknowledgments

We thank Anupam Gupta and László Lovász for useful discussions.

## A  Conditioning and Variance Reduction of Lasserre solutions

For any $S \subseteq [n]$, let $[k]^S$ be the set of all possible labelings of $S$. Recall the vectors $x_S(f)$ defined in (14). Let $[k]^\emptyset = \{\top\}$ where $\top$ denotes the (only) labeling of empty set with $x_\emptyset(\top) = x_\emptyset$ being some constant unit vector.

**Definition 27.** *Given $S \subseteq [n]$ and $f \in [k]^S$ with $x_S(f) \neq 0$, we define the vectors conditioned on $f$ as the following. For any $A \subseteq [n]$ and $g \in [k]^A$, the vector $x_{A|f}(g)$ is given by:*

$$x_{A|f}(g) \triangleq \frac{x_{S \cup A}(f \circ g)}{\|x_S(f)\|}.$$

Formally the conditional vectors $x_{A|f}(g)$ correspond to relaxations of respective indicator variables. Thus such vectors behave exactly in the same way with non-conditional vectors. Some of these properties are given in the following easy claim, whose proof we skip. For $g \in [k]^A$ and $h \in [k]^B$ that are consistent on $A \cap B$, we denote by $g \circ h \in [k]^{A \cup B}$ the labeling that restricts to $g$ (resp. $h$) on $A$ (resp. $B$).

**Claim 28.** *For any $f \in [k]^S$ with $x_S(f) \neq 0$, the following are true:*

(a) $x_{\emptyset|f}(\top) = x_{S|f}(f)$ and $\|x_{S|f}(f)\|^2 = 1$.

(b) *For any $g \in [k]^A$ and $h \in [k]^B$, we have $\langle x_{A|f}(g), x_{B|f}(h) \rangle = \|x_{A \cup B|f}(g \circ h)\|^2$ if $g, h$ are consistent on $A \cap B$ and $0$ otherwise.*

(c) For any $g \in [k]^A$, we have $x_A(g) = \sum_f \|x_S(f)\| x_{A|f}(g)$ so that
$\|x_A(g)\|^2 = \sum_f \|x_S(f)\|^2 \|x_{A|f}(g)\|^2$.

(d) For any $g \in [k]^A$, $\|x_{A|f}(g)\|^2 = \frac{\langle x_S(f), x_A(g) \rangle}{\|x_S(f)\|^2}$.

(e) For any $g \in [k]^A$ and $h \in [k]^B$,
$$x_{B|f,g}(h) = \frac{x_{A \cup B|f}(g \circ h)}{\|x_{A|f}(g)\|}.$$

*Proof.* Items a to d are easy. For item e, by definition:
$$x_{B|f,g}(h) = \frac{x_{S \cup A \cup B}(f \circ g \circ h)}{\|x_{S \cup A}(f \circ g)\|} = \frac{x_{A \cup B|f}(g \circ h) \cdot \|x_S(f)\|}{\|x_{A|f}(g)\| \cdot \|x_S(f)\|}. \qquad \square$$

Assume that some labeling $f_0 \in [k]^{S_0}$ to $S_0$ has been fixed, and we further sample a labeling $f$ to $S$ with probability $\|x_{S|f_0}(f)\|^2$ (i.e., from the conditional probability distribution of labelings to $S$ given labeling $f_0$ to $S_0$). The following defines a projection matrix which captures the effect of further conditioning according to the labeling to $S$. For a nonzero vector $v$, we denote by $\bar{v}$ the unit vector in the direction of $v$.

**Notation 29.** *Given $f_0 \in [k]^{S_0}$ and $S \subseteq [n]$, let*
$$\Pi_{S|f_0} \triangleq \sum_{f: x_{S|f_0}(f) \neq 0} \overline{x_{S|f_0}(f)} \cdot \overline{x_{S|f_0}(f)}^T.$$

*Similarly let $\Pi_S^\perp \triangleq I - \Pi_S$ where $I$ is the identity matrix of the appropriate dimension.*

We will now relate properties of the conditional probability distribution arising from partial labelings to the above projection matrix. First we will define the random variables corresponding to each indicator function with matching moments:

**Definition 30.** *Given $f \in [k]^S$, for all $g \in [k]^A, h \in [k]^B$, let $\mathcal{X}_{A|f}(g)$ and $\mathcal{X}_{B|f}(h)$ be two random variables over $\{0, 1\}$ such that:*
$$\mathrm{Prob}[\mathcal{X}_{A|f}(g) = 1 \wedge \mathcal{X}_{B|f}(h)] = \langle x_{A|f}(g), x_{B|f}(h) \rangle.$$

The above definition suggests a very simple rounding scheme: Choose a label for each variable based on this probability. In fact, all rounding algorithms we can handle in our framework carry this trait. One way to measure how far we can go with only these probabilities is to look at their variance:

**Claim 31.** $\mathrm{Var}(\mathcal{X}_{A|f}(g)) = \|x_{A|f}(g)\|^2 - \|x_{A|f}(g)\|^4 = \|x_{\emptyset|f}^\perp x_{A|f}(g)\|^2$.

*Proof.* Since $\mathcal{X}_{A|f}(g)$ is a random variable over $\{0, 1\}$,
$$\mathrm{Var}(\mathcal{X}_{A|f}(g)) = \mathbb{E}[\mathcal{X}_{A|f}(g)](1 - \mathbb{E}[\mathcal{X}_{A|f}(g)]) = \|x_{A|f}(g)\|^2 - \langle x_{\emptyset|f}, x_{A|f}(g) \rangle^2 = \|x_{\emptyset|f}^\perp x_{A|f}(g)\|^2. \quad \square$$

**Claim 32.** $\mathrm{Cov}(\mathcal{X}_{A|f}(g), \mathcal{X}_{B|f}(h)) = \langle x_{\emptyset|f}^\perp x_{A|f}(g), x_{\emptyset|f}^\perp x_{B|f}(h) \rangle$.

*Proof.* Since $\mathbb{E}[\mathcal{X}_{A|f}(g)\mathcal{X}_{B|f}(h)] = \langle x_{A|f}(g), x_{B|f}(h) \rangle$, we can express $\text{Cov}(\mathcal{X}_{A|f}(g), \mathcal{X}_{B|f}(h))$ as:

$$= \langle x_{A|f}(g), x_{B|f}(h) \rangle - \langle x_{\emptyset|f}, x_{A|f}(g) \rangle \langle x_{\emptyset|f}, x_{B|f}(h) \rangle = \langle x^{\perp}_{\emptyset|f} x_{A|f}(g), x^{\perp}_{\emptyset|f} x_{B|f}(h) \rangle. \qquad \square$$

The following was first observed in [GS11], and enables controlling probabilistic quantities in terms of a geometric quantity, the projection distance to certain subspaces. In particular, it says that if we can somehow choose $S$ and $f_0$ in such a way that span of $\Pi_{S|f_0}$ is very close to the vectors $x_{u|f_0}$, then the variance will be small.

**Lemma 33.** *Given $f_0 \in [k]^{S_0}$, subsets $S, A \subseteq [n]$, and $g \in [k]^A$, we have*

$$\mathbb{E}_{f \sim \|x_{S|f_0}(f)\|^2}\left[\text{Var}(\mathcal{X}_{A|f,f_0}(g))\right] = \|\Pi^{\perp}_{S|f_0} x_{A|f_0}(g)\|^2.$$

*Proof.* Using Claim 31, we see that:

$$\mathbb{E}_{f \sim \|x_{S|f_0}(f)\|^2}\left[\text{Var}(\mathcal{X}_{A|f,f_0}(g))\right] = \sum_f \|x_{S|f_0}(f)\|^2 \left(\|x_{A|f,f_0}(g)\|^2 - \|x_{A|f,f_0}(g)\|^4\right)$$

$$= \|x_{A|f_0}(g)\|^2 - \sum_f \|x_{S|f_0}(f)\|^2 \|x_{A|f,f_0}(g)\|^4 \qquad \text{(using Claim 28 (c))}$$

$$= \|x_{A|f_0}(g)\|^2 - \sum_f \|x_{S|f_0}(f)\|^2 \langle x_{\emptyset|f,f_0}, x_{A|f,f_0}(g) \rangle^2 \qquad \text{(using Claim 28 (b))}$$

$$= \|x_{A|f_0}(g)\|^2 - \sum_f \langle x_{S|f_0}(f), x_{A|f,f_0}(g) \rangle^2 \qquad \text{(using Definition 27)}$$

$$= \|x_{A|f_0}(g)\|^2 - \sum_{f:x_{S|f_0}(f) \neq 0} \left\langle x_{S|f_0}(f), \frac{x_{S \cup A|f_0}(f \circ g)}{\|x_{S|f_0}(f)\|} \right\rangle^2 \qquad \text{(using Claim 28 (e))}$$

$$= \|x_{A|f_0}(g)\|^2 - \sum_{f:x_{S|f_0}(f) \neq 0} \frac{1}{\|x_{S|f_0}(f)\|^2} \langle x_{S|f_0}(f), x_{A|f_0}(g) \rangle^2 \qquad \text{(using Claim 28 (b))}$$

$$= \|x_{A|f_0}(g)\|^2 - \sum_f \langle \overline{x_{S|f_0}(f)}, x_{A|f_0}(g) \rangle^2$$

$$= \|\Pi^{\perp}_{S|f_0} x_{A|f_0}(g)\|^2 \qquad \text{(using Notation 29).} \qquad \square$$

## B  Other Rounding Algorithms

In this section, we will show how the partial coloring algorithm from [AG11] and 2-CSP algorithm from [BRS11] fit into our framework. The main difficulty is that both these algorithms are adaptive. In particular, a naive adaptation will have $\ell = \Omega(r)$ which is quite undesirable for our faster solver. We can easily get around this difficulty by replacing the adaptive seed selection procedure with a suitable version of Algorithm 4.

### B.1  Partial Coloring of 3-Colorable Graphs

**Theorem 34.** *Given a 3-colorable d-regular graph $G$ on $n$ nodes, positive real $1 > \varepsilon > 0$ and positive integer $r$, suppose its $r^{th}$ largest eigenvalue of normalized Laplacian matrix, $\lambda_{n-r}$, satisfies*

$$\lambda_{n-r} \leqslant \frac{4 - \delta}{3}$$

for some positive real $\delta > 0$. Then, for the choice of $r' = O(r/\delta\varepsilon)$, Algorithm 5 followed by the rounding algorithm as described in [AG11] will output a partial coloring which colors at least $(1-\varepsilon)\frac{\delta}{2+\delta}n$ nodes. Furthermore this algorithm can be implemented in time $\text{poly}(n)2^{O(r/\delta\varepsilon)}$ using the faster solver framework.

The main advantage of our seed selection procedure (which enables the speed-up using our faster solver) is that we pick $r'$ nodes all at once, instead of picking them one-by-one in $r'$ steps as in [AG11]. We have the following as an immediate corollary of Theorem 34:

**Corollary 35.** *Given a 3-colorable $d$-regular graph $G$, for any positive integer $r$ with $\lambda_{n-r} \leq \frac{10}{9} - \Omega(1)$, we can find a partial coloring on $\frac{n}{4}$ nodes and an independent set of size at least $\frac{n}{12}$ in time $\text{poly}(n)2^{O(r)}$.*

Before we begin the proof of Theorem 34, we will state some simple claims. As the method applies for $k$-colorable graphs with different parameters, below for clarity we first use $k$ for the number of colors, and then later set $k = 3$.

**Claim 36.** *For any edge $(u,v)$ of $G$,*

$$\frac{1}{2}\|X_u + X_v\|^2 \leq 1 - \frac{2}{k}.$$

*In particular, if we use $A$ to denote the normalized adjacency matrix of $G$, then:*

$$\text{Tr}\left[X^T X(I + A)\right] \leq n\left(1 - \frac{2}{k}\right).$$

*Proof.*

$$\begin{aligned}
\frac{1}{2}\|X_u + X_v\|^2 &= \frac{1}{2}\sum_{i\in[k]}\left(\|x_\emptyset^\perp x_u(i)\|^2 + \|x_\emptyset^\perp x_v(i)\|^2 + 2\langle x_\emptyset^\perp x_u(i), x_\emptyset^\perp x_v(i)\rangle\right) \\
&= \frac{1}{2}\sum_{i\in[k]}\Big(\|x_u(i)\|^2 - \|x_u(i)\|^4 + \|x_v(i)\|^2 - \|x_v(i)\|^4 \\
&\quad + 2\langle x_u(i), x_v(i)\rangle - 2\langle x_\emptyset, x_u(i)\rangle\langle x_\emptyset, x_v(i)\rangle\|x_\emptyset\|^2\Big) \\
&= 1 - \frac{1}{2}\sum_{i\in[k]}\left(\|x_u(i)\|^4 + \|x_v(i)\|^4 + 2\|x_u(i)\|^2\|x_v(i)\|^2\right) \text{ (using } \langle x_u(i), x_v(i)\rangle = 0) \\
&= 1 - \frac{1}{2}\sum_{i\in[k]}\left(\|x_u(i)\|^2 + \|x_v(i)\|^2\right)^2.
\end{aligned}$$

At this point, observe that $\sum_{i\in[k]}\left(\|x_u(i)^2 + \|x_v(i)\|^2\right)^2$ is a convex function on $\|x_u(i)\|^2$ and $\|x_v(j)\|^2$'s. Since $\sum_i \|x_u(i)\|^2 = \sum_j \|x_v(j)\|^2 = 1$, it is minimized when $\|x_u(i)\|^2 = \|x_v(j)\|^2 = \frac{1}{k}$. Substituting this into the above expression, we see that:

$$\frac{1}{2}\|X_u + X_v\|^2 \leq 1 - \frac{k}{2}\left(\frac{2}{k}\right)^2 = 1 - \frac{2}{k}.$$

For the final part, observe that:

$$\text{Tr}\left[X^T X(I+A)\right] = \frac{1}{d}\sum_{\{u,v\}\in E(G)}\|X_u + X_v\|^2 \leq \frac{2|E(G)|}{d}\left(1 - \frac{2}{k}\right) = n\left(1 - \frac{2}{k}\right). \qquad \square$$

**Claim 37.** *Given a graph $G$ and positive integer $r$, for $\lambda_r$ being the $r^{th}$ smallest eigenvalue of corresponding normalized graph Laplacian matrix, the following holds:*

$$\sum_{j \geq r} \sigma_j(X^T X) \leq n \frac{1 - 2/k}{2 - \lambda_r}.$$

*Proof.* Follows from using the upper bound from Claim 36 on inequality:

$$\sum_{j \geq r} \sigma_j(X^T X) \leq \frac{1}{\lambda_r} \text{Tr}\left[X^T X (I + A)\right].$$

□

**Claim 38.** *Assume $u$ is uncolored. Then:*

$$\sum_i \|x_{\emptyset|f}^\perp x_{u|f(i)}\|^2 \geq \frac{1}{2}.$$

*Proof.* Note that $\|x_{\emptyset|f}^\perp x_{u|f}(i)\|^2 = \|x_{u|f}(i)\|^2(1 - \|x_{u|f}(i)\|^2)$. If $u$ is uncolored, then $1 - \|x_{u|f}(i)\|^2 \geq \frac{1}{2}$ for all $i \in [k]$[5], in which case we have:

$$\sum_i \|x_{\emptyset|f}^\perp x_{u|f(i)}\|^2 \geq \frac{1}{2} \sum_i \|x_{u|f}(i)\|^2 = \frac{1}{2}.$$

□

For a subset $S$ of vertices of $G$, we denote by $X_S^\perp$ the projection operator onto the orthogonal complement of $\text{span}\{X_u \mid u \in S\}$.

**Lemma 39.** *For coloring $f$ to a subset $S$ sampled with probability $\|x_S(f)\|^2$:*

$$\mathbb{E}_f \left[\sum_i \|x_{\emptyset|f}^\perp x_{u|f(i)}\|^2\right] \leq \|X_S^\perp X_u\|^2.$$

*Proof.* From Lemma 33, we know that $\mathbb{E}_f \left[\sum_i \|x_{\emptyset|f}^\perp x_{u|f}(i)\|^2\right] = \sum_i \left\|\Pi_S^\perp x_u(i)\right\|^2 \leq \|X_S^\perp X_u\|^2$. The final inequality follows from the same arguments as in [GS11]. □

*Proof of Theorem 34.* Let $\delta' = \frac{1}{2}\delta$ and $\varepsilon' = \varepsilon\delta'$. By [GS12b], we know that volume sampling of $r' = O(r/\varepsilon')$ columns yields

$$\sum_u \|X_S^\perp X_u\|^2 \leq (1 + \varepsilon) \sum_{j \geq r} \sigma_j(X^T X) \leq n(1 + \varepsilon') \frac{1 - 2/k}{2 - \lambda_r}.$$

Using Markov inequality, the fraction of uncolored nodes is bounded by:

$$\leq 2n(1 + \varepsilon) \frac{1 - 2/k}{2 - \lambda_r} = \frac{2(1 + \varepsilon')}{3(2 - \lambda_r)} n \quad \text{(for } k = 3\text{)}.$$

For $\lambda_r \leq \frac{4}{3} - \frac{2}{3}\delta'$, this expression becomes $\frac{1+\varepsilon'}{1+\delta'}n$, which implies

$$\mathbb{E}\,[\text{fraction of colored nodes}] \geq 1 - \frac{1 + \varepsilon'}{1 + \delta'} = \frac{\delta' - \varepsilon'}{1 + \delta'} = \frac{\delta'}{1 + \delta'}(1 - \varepsilon) = \frac{\delta/2}{1 + \delta/2}(1 - \varepsilon) = \frac{\delta}{2 + \delta}(1 - \varepsilon).$$

To prove that the coloring output is legal, notice that for any pair of adjacent nodes $(u, v) \in E(G)$, both $\|x_{u|f}(i)\|^2$ and $\|x_{v|f}(i)\|^2$ cannot be larger than $1/2$ both at the same time. □

---

[5]This follows from the threshold rounding algorithm used in [AG11] for coloring, which colors $u$ with color $i$ if $\|x_{u|f}(i)\|^2 > 1/2$.

## B.2 Approximating 2-CSPs

Given a 2-CSP problem on variables $[n]$ and labels $[k]$, let $G$ be its constraint graph. For convenience, we assume $G$ is regular; however all our bounds still hold when $G$ is non-regular. We use $A$ to denote $G$'s normalized adjacency matrix and $\lambda_i$ to denote the $i^{th}$ smallest eigenvalue of $G$'s normalized Laplacian matrix. Finally we will use $uv \sim G$ to denote sampling a constraint with probability proportional to the weight of constraint between $u$ and $v$.

**Embedding.** Consider the embedding used in Lemma 5.3 of [BRS11] which is used to convert $k$ vectors $x_u(i)$ into a single vector. Given a partial assignment $f \in [k]^S$ and $u \in [n]$ with $(x_{u|f}(i))_{i \in [k]} \subset \mathbb{R}^{[m]}$, we define $X_u(f)$ as the following vector.

$$X_u(f) \triangleq \frac{1}{\sqrt{k}} \sum_j \frac{(x_{\emptyset|f}^\perp x_{u|f}(j))^{\otimes 2}}{\|x_{\emptyset|f}^\perp x_{u|f}(j)\|} \tag{15}$$

**Seed Selection and Rounding.** We will give only an overview of the seed selection procedure. In [BRS11], assuming some lower bound on $\lambda_r$ (in terms of $\varepsilon, k$, where $\varepsilon$ is the additive approximation error), a seed set of $r \cdot \text{poly}(k/\varepsilon)$ vertices will be picked in as many iterations, one vertex at a time. We modify the seed selection to involve fewer adaptive stages, with $\ell = O(k^2/\varepsilon^2)$ stages each picking $O(r/\varepsilon)$ vertices each. Plugging into our general solver then gives a runtime improvement as before.

At $i^{th}$ level, we choose a seed set of size $O(r/\varepsilon)$, $S_i$, from the matrix $X(f_i) = [X_u(f_i)]_{u \in [n]}$ where $X_u$'s are defined in eq. (15). After choosing seed set $S_i$, we sample an assignment $g_i \in [k]^{S_i}$ (conditioned on $f_{i-1}$) that satisfies

$$\delta_{f_{i-1}, g_i} \leqslant \mathbb{E}_{g \sim \|x_{S|f_{i-1}}(g)\|^2} \left[ \delta_{f_{i-1}, g} \right]$$

where $\delta_f$ is defined in eq. (17) and set $f_i \leftarrow f_{i-1} \circ g_i$. We repeat the seed selection procedure as long as $\varepsilon_{f_i} > \varepsilon$ where $\varepsilon_f$ is as defined in eq. (16).

The rounding procedure remains the same — independent labeling for each CSP variable from the respective conditional distributions. Formally, for each variable $u \in [n]$, we choose a label $i \in [k]$ with probability $\|x_{u|f_i}(i)\|^2$ independently at random. In Theorem 44, we will show that $\ell = O\left(\frac{k^2}{\varepsilon^2}\right)$, i.e. seed selection will terminate after choosing at most $\ell$ sets.

**Analysis.** Let us begin by defining the quantity

$$\varepsilon_f \triangleq \mathbb{E}_{uv \sim G} \sum_{(i,j) \in [k]^2} \left| \mathbb{E}\left[\mathcal{X}_{uv|f}(ij)\right] - \mathbb{E}\left[\mathcal{X}_{u|f}(i)\right] \mathbb{E}\left[\mathcal{X}_{v|f}(j)\right] \right|$$
$$= \mathbb{E}_{uv \sim G} \sum_{i,j} \left| \text{Cov}\left[\mathcal{X}_{u|f}(i), \mathcal{X}_{v|f}(j)\right] \right| = \mathbb{E}_{uv \sim G} \sum_{i,j} \left| \langle x_{\emptyset|f}^\perp x_{u|f}(i), x_{\emptyset|f}^\perp x_{u|f}(j) \rangle \right|. \tag{16}$$

As shown in [BRS11], the above gives an upper bound on the expected extra fraction of unsatisfied constraints in the rounded solution compared to the Lasserre SDP optimum (when performing rounding after conditioning on assignment $f$). Therefore, when $\varepsilon_f \leqslant \varepsilon$, we get an additive $\varepsilon$-error approximation. Our goal is prove (which we will do in Theorem 44) that for $\ell \leqslant \tilde{O}(k/\varepsilon)$, we must have $\varepsilon_{f_\ell} \leqslant \varepsilon$.

If we define the quantity $\delta_f$ measuring the expected total variances of each $\mathcal{X}_{u|f}(i)$ as

$$\delta_f \triangleq \mathbb{E}_u \sum_{i \in [k]} \text{Var}\left[\mathcal{X}_{u|f}(i)\right] = \mathbb{E}_u \sum_{i \in [k]} \|x^\perp_{\emptyset|f} x_{u|f}(i)\|^2 , \quad (17)$$

then it is easy to see that $\varepsilon_f \leqslant k\delta_f$ by Cauchy-Schwarz.

We will first relate eq. (16) to the inner products of the embedded vectors $X_u(f)$.

**Claim 40.** $\mathbb{E}_{uv \sim G}[\langle X_u(f), X_v(f) \rangle] \geqslant \left(\frac{\varepsilon_f}{k}\right)^2$.

*Proof.* We have

$$k\langle X_u(f), X_v(f) \rangle = \sum_{ij} \frac{\langle x^\perp_{\emptyset|f} x_{u|f}(i), x^\perp_{\emptyset|f} x_{v|f}(j) \rangle^2}{\|x^\perp_{\emptyset|f} x_{u|f}(i)\| \|x^\perp_{\emptyset|f} x_{v|f}(j)\|} \geqslant \frac{\left(\sum_{ij} \left|\langle x^\perp_{\emptyset|f} x_{u|f}(i), x^\perp_{\emptyset|f} x_{v|f}(j) \rangle\right|\right)^2}{\sum_{ij} \|x^\perp_{\emptyset|f} x_{u|f}(i)\| \|x^\perp_{\emptyset|f} x_{v|f}(j)\|} \quad (18)$$

where the second step uses Cauchy Schwarz. Since

$$\sum_i \|x^\perp_{\emptyset|f} x_{u|f}(i)\| \leqslant \sqrt{k} \left(\sum_i \|x^\perp_{\emptyset|f} x_{u|f}(i)\|^2\right)^{1/2} \leqslant \sqrt{k} \left(\sum_i \|x_{u|f}(i)\|^2\right)^{1/2} = \sqrt{k}$$

the expected value of the above lower bound (18) for $uv \sim G$ is at least $\varepsilon_f^2/k$. $\square$

We now upper bound the lengths of the embedded vectors.

**Claim 41.** $\|X_u(f)\|^2 \leqslant \sum_{i \in [k]} \left\|x^\perp_{\emptyset|f} x_{u|f}(i)\right\|^2$. In particular, $\mathbb{E}_u \|X_u(f)\|^2 \leqslant \delta_f$.

*Proof.* $\frac{1}{k}\|X_u(f)\|^2 = \mathbb{E}_{ij} \frac{\langle x^\perp_{\emptyset|f} x_{u|f}(i), x^\perp_{\emptyset|f} x_{u|f}(j) \rangle^2}{\|x^\perp_{\emptyset|f} x_{u|f}(i)\| \|x^\perp_{\emptyset|f} x_{u|f}(j)\|} \leqslant \left(\mathbb{E}_i \left\|x^\perp_{\emptyset|f} x_{u|f}(i)\right\|\right)^2 \leqslant \mathbb{E}_i \left\|x^\perp_{\emptyset|f} x_{u|f}(i)\right\|^2$. $\square$

Now for fixed $f$ we will upper bound the expected value of $\delta_{f,g}$ over $g \sim \|x_{S|f}(g)\|^2$ in terms of the projection distance of the embedded vectors from the subspace spanned by $X_v(f)$ for $v \in S$. (Below, $X_S(f)^\perp$ denotes the projection onto the orthogonal complement of $\text{span}\{X_v(f) \mid v \in S\}$.)

**Claim 42.** $\mathbb{E}_{g \sim \|x_{S|f}(g)\|^2}[\delta_{f,g}] \leqslant \mathbb{E}_{u \sim G}\left[\left\|X_S^\perp(f) X_u(f)\right\|^2\right]$.

*Proof.* By Lemma 33, we know that $\mathbb{E}_g\left[\left\|x^\perp_{\emptyset|g,f} x_{u|g,f}(i)\right\|^2\right] = \left\|\Pi^\perp_{S|f} x_{u|f}(i)\right\|^2$. Since $x_{\emptyset|f}$ is in the span of $\Pi_{S|f}$ by Claim 28(c), $\Pi^\perp_{S|f} x_{u|f}(i) = \Pi^\perp_{S|f} x^\perp_{\emptyset|f} x_{u|f}(i)$. Similarly for any $v \in S$ and $j \in [k]$, the vector $x^\perp_{\emptyset|f} x_{v|f}(j)$ is in the span of $\Pi_{S|f}$. By using the same arguments from [GS11][6], namely the embedding used here preserves linearity, we obtain $\sum_i \left\|\Pi^\perp_{S|f} x^\perp_{\emptyset|f} x_{u|f}(i)\right\|^2 \leqslant \left\|X_S^\perp(f) X_u(f)\right\|^2$. Taking expectation over $u$ completes the proof. $\square$

Using the above, we can prove the main claim about the seed selection procedure, namely that, assuming $\lambda_r$ is close enough to 1, the expected variance $\delta_f$ can be reduced by a geometric factor by conditioning on the assignment to a further $O(r/\varepsilon)$ nodes.

---
[6]See Claim 24 in the full version of [GS11].

**Lemma 43.** *Given $f \in [k]^{S_0}$, positive real $\varepsilon > 0$ and positive integer $r$ with $\lambda_{r+1} \geq 1 - \frac{\varepsilon^2}{2k^2}$, if $\varepsilon_f \geq \varepsilon$ then there exists a set of $O(rk^2/\varepsilon^2)$-columns of $X(f)$, $S$ and $g \in [k]^S$ such that $x_{S|f}(g) \neq 0$ and:*

$$\delta_{f,g} \leq \delta_f - \Omega\left(\frac{\varepsilon^2}{k^2}\right). \tag{19}$$

*Furthermore such $S$ and $g$ can be found in $\text{poly}(n)k^{O(r/\varepsilon)}$ time using volume sampling [DR10, GS12b] to find $S$ and then enumerating all $g$'s.*

*Proof.* Let $\rho \triangleq \varepsilon/k$, and $\mu \triangleq \mathbb{E}_u \|X_u\|^2$ where for notational convenience we suppress the dependence on $f$ and denote $X_u(f)$ by $X_u$. Observe that

$$\frac{1}{n} \text{Tr}\left[X^T X A\right] = \mathbb{E}_{uv \sim G}\langle X_u, X_v \rangle \geq (\varepsilon_f/k)^2 \geq \rho^2$$

by Claim 40. This implies $\frac{1}{n}\text{Tr}\left[X^T X L\right] \leq \mathbb{E}_u \|X_u\|^2 - \rho^2 = \mu - \rho^2$. From Lemma 30 of [GS11], we know that volume sampling $O(r/\rho^2)$ columns from $X$ yields a set $S$ for which:

$$\mathbb{E}_u \|X_S^\perp X_u\|^2 \leq \left(1 + O(\rho^2)\right) \frac{(1/n)\text{Tr}\left[X^T X L\right]}{1 - \max(1 - \lambda_{r+1}, 0)} \leq \left(1 + O(\rho^2)\right) \frac{\mu - \rho^2}{1 - \frac{\rho^2}{2}}$$

Since $\rho \leq 1$, we have $(1 - \rho^2/2)^{-1} \leq \left(1 + \frac{3}{4}\rho^2\right)$:

$$\leq \left(1 + O(\rho^2)\right)(\mu - \rho^2)\left(1 + \frac{3}{4}\rho^2\right) \leq \left(1 + O(\rho^2)\right)\left(\mu - \frac{\rho^2}{4}\right)$$
$$\leq \mu - \Omega\left(\rho^2\right)$$
$$\leq \delta_f - \Omega\left(\rho^2\right) \qquad \text{(by Claim 41)}.$$

By Claim 42, $\mathbb{E}_g[\delta_{f,g}] \leq \mathbb{E}_u \|X_S^\perp X_u\|^2$, which means there exists $g$ for which $\delta_{f,g} \leq \delta_f - \Omega\left(\frac{\varepsilon^2}{k^2}\right)$. □

We put together everything in the following theorem.

**Theorem 44.** *For $\ell = O\left(\frac{k^2}{\varepsilon^2}\right)$, seed selection procedure will output a partial assignment $f_\ell$ with $\varepsilon_{f_\ell} \leq \varepsilon$.*

*Proof.* Suppose $\varepsilon_{f_i} > \varepsilon$ for all $i \leq \ell$. Then by Lemma 43, for each $i \leq \ell$:

$$0 \leq \delta_{f_i} \leq \delta_{f_0} - i\Omega\left(\frac{\varepsilon^2}{k^2}\right) \leq 1 - \Omega\left(\frac{i\varepsilon^2}{k^2}\right) \implies \delta_{f_\ell} < 0,$$

which is a contradiction. □